\documentclass{aa}  

\usepackage{natbib,twoopt}
\usepackage[breaklinks=true]{hyperref} 
\bibpunct{(}{)}{;}{a}{}{,}             
\makeatletter
  \newcommandtwoopt{\citeads}[3][][]{\href{http://adsabs.harvard.edu/abs/#3}%
    {\def\hyper@linkstart##1##2{}%
     \let\hyper@linkend\@empty\citealp[#1][#2]{#3}}}
  \newcommandtwoopt{\citepads}[3][][]{\href{http://adsabs.harvard.edu/abs/#3}%
    {\def\hyper@linkstart##1##2{}%
     \let\hyper@linkend\@empty\citep[#1][#2]{#3}}}
  \newcommandtwoopt{\citetads}[3][][]{\href{http://adsabs.harvard.edu/abs/#3}%
    {\def\hyper@linkstart##1##2{}%
     \let\hyper@linkend\@empty\citet[#1][#2]{#3}}}
  \newcommandtwoopt{\citeyearads}[3][][]%
    {\href{http://adsabs.harvard.edu/abs/#3}
    {\def\hyper@linkstart##1##2{}%
     \let\hyper@linkend\@empty\citeyear[#1][#2]{#3}}}
\makeatother

\usepackage{longtable}
\usepackage{graphicx}
\hypersetup{
  colorlinks = true,
  urlcolor = purple,
  citecolor = purple,
  linkcolor = purple,
}

\usepackage{txfonts}
\usepackage{xcolor}
\begin{document}

   \title{Mining archival data from wide-field astronomical surveys in search of near-Earth objects}
   \titlerunning{Mining archival data from wide-field astronomical surveys in search of NEOs}
   \authorrunning{T. Saifollahi et al.}


   
   
   \author{Teymoor Saifollahi
          \inst{1},
          Gijs Verdoes Kleijn\inst{1}, Rees Williams\inst{1}, Marco Micheli\inst{2}, Toni Santana-Ros\inst{3,4}, \\Ewout Helmich\inst{1}, Detlef Koschny\inst{5}, Luca Conversi\inst{2,6}
          }

   \institute{\inst{1} Kapteyn Astronomical Institute, University of Groningen, PO Box 800, 9700 AV Groningen, The Netherlands\\
   \inst{2} ESA PDO NEO Coordination Centre, Largo Galileo Galilei, 1, 00044 Frascati, Italy\\
   \inst{3} {Instituto de Física Aplicada a las Ciencias y las Tecnolog\'{\i}as, Universidad de Alicante, San Vicente del Raspeig, 03080, Alicante, Spain}\\
    \inst{4} {Institut de Ci\`{e}ncies del Cosmos (ICCUB), Universitat de Barcelona (IEEC-UB), Carrer de Mart\'{\i} i Franqu\`{e}s, 1, 08028 Barcelona, Spain}\\
    \inst{5} ESA/ESTEC, Keplerlaan 1, 2200 XH Noordwijk, The Netherlands\\
    \inst{5} TU Munich, Boltzmannstr. 15, 85748 Garching, Germany\\
    \inst{6} European Space Agency / ESRIN, Largo Galileo Galilei, 1, 00044 Frascati, Italy
    }
   \date{Received ...; accepted ...}
   

 
  \abstract
  {Increasing our knowledge of the orbits and compositions of Near-Earth Objects (NEOs) is important for a better understanding of the evolution of the Solar System and of life. The detection of serendipitous NEO appearances among the millions of archived exposures from large astronomical imaging surveys can provide a contribution which is complementary to NEO surveys.}
  {Using the \textsc{AstroWISE} information system, this work aims to assess the detectability rate, the achieved recovery rate and the quality of astrometry when data mining European Southern Observatory (ESO)’s archive for the OmegaCAM wide-field imager at the VLT Survey Telescope (VST).}
  {We developed an automatic pipeline that searches for the NEO appearances inside the \textsc{AstroWISE} environment. Throughout the recovery process, the pipeline uses several public web-tools (SSOIS, NEODyS, JPL Horizons) to identify possible images that overlap with the position of NEOs, and acquires information on the NEOs predicted position and other properties (e.g., magnitude, rate and direction of motion) at the time of observations. Considering these properties, the pipeline narrows down the search to potentially detectable NEOs, searches for streak-like objects across the images and finds a matching streak for the NEOs.}
  {We have recovered 196 appearances of NEOs from a set of 968 appearances predicted to be recoverable. It includes appearances for three NEOs which were on the impact risk list at that point. These appearances were well before their discovery. The subsequent risk assessment using the extracted astrometry removes these NEOs from the risk list. More generally, we estimate a detectability rate of $\sim$0.05 per NEO at a signal-to-noise ratio larger than 3 for NEOs in the OmegaCAM archive. Our automatic recovery rates are 40\% and 20\% for NEOs on the risk list and the full list, respectively. The achieved astrometric and photometric accuracy is on average 0.12\arcsec and 0.1\,mag.}
  {These results show the high potential of the archival imaging data of the ground-based wide-field surveys as useful instruments for the search, (p)recovery and characterization of NEOs. Highly automated approaches, as possible using \textsc{AstroWISE}, make this undertaking feasible.}

  \maketitle
%

\section{Introduction}

Minor bodies in planetary systems potentially have an intimate relationship with life. Minor bodies might provide complex molecules to planets to facilitate the formation of life (e.g., \citealp{Oba22} while impacts later on can play a significant role in its evolution through climate changes and mass extinctions (e.g., \citealp{Alvarez}). 
Deepening our understanding on this potential relationship is a high priority on both the European science agenda (e.g., Cosmic Voyage 2050\footnote{\href{https://www.cosmos.esa.int/web/voyage-2050}{https://www.cosmos.esa.int/web/voyage-2050}}) and the US science agenda (e.g., "Origins, Worlds, and Life: A Decadal Strategy for Planetary Science and Astrobiology 2023-2032\footnote{\href{https://www.nationalacademies.org/our-work/planetary-science-and-astrobiology-decadal-survey-2023-2032\#sectionPublications}{https://www.nationalacademies.org/our-work/planetary-science-and-astrobiology-decadal-survey-2023-2032\#sectionPublications}}). 
Near-Earth Objects are a reservoir that interacts surely with Earth’s biosphere and represents physically the closest instance of minor bodies that are at our disposal. NEOs are asteroids or comets whose perihelion occurs at less than 1.3 astronomical units (au), meaning that close approaches with the Earth might occur at some point. The size of these objects ranges from meters to tens of kilometres. 
Currently, more than 30~000 near-Earth objects (NEOs) are catalogued in the Minor Planet Center (MPC\footnote{\url{https://www.minorplanetcenter.net/}}) and the discovery rate has reached the order of thousands per year. While we are certain that the vast majority of the largest NEOs have been already discovered, very little is known about the majority: the ones in the range of a few meters up to about 150 meters.

With NEOs, the minor body $\leftrightarrow$ life relationship can be approached from a scientific perspective: understanding the role of minor bodies in the origin and long-term evolution of life in a planetary system, such as our own Solar System. For example, NEOs might deliver the meteorites that provide the "starter set" for life in the form of prebiotic molecules (\citealp{Oba22}). Furthermore, NEO compositions can put constraints on competing formation scenarios for rocky planets like Earth \citep{burkhardt21}. Thus the characterization of the orbits and physical composition of Near-Earth Objects is valuable to advance our scientific understanding of the minor body $\leftrightarrow$ life relationship.

The minor bodies $\leftrightarrow$ life relationship can also be approached from a societal perspective: the impact hazard poses a significant threat to technological civilizations on a planet (\citealp{Borovicka13}, \citealp{Popova13}, \citealp{Brown13} ) and calls for planetary defence strategies. The Tunguska and Chelyabinsk impact are recent reminders. Advanced technological civilizations on a planet can also use their neighbouring minor bodies to mine (rare) chemical elements and compounds \citep{hein20}, possibly as a foraging stop in space exploration \citep{davis93}. In conclusion, characterization of the orbits and physical composition of Near-Earth Objects is valuable both to advance both our scientific understanding of the formation and evolution of planetary systems and life and to serve societal goals related to planetary defence and space-based technological infrastructures. 

Small NEOs are only observable from the Earth during their brief close approaches with our planet. Every day, observatories such as Pan-STARRS (\citealp{pan}), Catalina \citep{christensen19}, ATLAS (\citealp{atlas2}) and the Zwicky Transient Facility (ZTF, \citealp{ztf}) are discovering new NEO candidates which need observational follow-up for their confirmation. It is not always easy to provide follow-up observations of these objects, since their changing observational conditions, usually linked to a growing position uncertainty, make them challenging objects for most telescopes. Particularly challenging are the observations of some subgroups of NEOs such as the Interior Earth Objects (IEOs, \citealp{Sheppard22}) or Earth companions such as the Trojans (see for instance \citealp{Santana-Ros222022}). Once the close approach with the Earth ends, it is impossible to gather any further data on the NEO, until the next close approach happens, which sometimes might take several years. The only way to get any additional information about these bodies once their observational period is over is to rely on data mining. Therefore, it is important to systematically mine and monitor new archival data for the appearance of these objects. This can be done using serendipitous discovery in archival observations not dedicated Near-Earth Objects. For example, the Dark Energy Survey (DES, \citealp{des,des-sso,des-sso2}), the Kilo-Degree Survey (KiDS, \citealp{kids-sso}) and the VISTA Hemisphere Survey (VHS, \citealp{vhs,vhs2}) have been used to search for minor bodies, including NEOs. Similarly, upcoming wide-area surveys such as ESA's \textit{Euclid} mission (\citealp{euclid-sso,streakdet2}) and the Legacy Survey of Space and Time (\citealp{lsst-sso2,lsst-sso}) and their combination \citep{guy22} will have a role in NEO astrometric and photometric observations. These astronomical surveys have complementary data, being often deeper and surveying more away from the ecliptic compared to NEO-dedicated surveys. 
In particular, NEO \textit{precovery} $-$ detection of a known NEO in an observing dataset prior to its discovery $-$ provides information about the NEO's motion at an earlier epoch at a likely complementary location in its orbit. This is because the orbit uncertainty depends on the fraction of the orbital arc that is covered during discovery. (P)recovering one or more points far away from the discovery and immediate follow-up observations can significantly improve the accuracy of the orbital parameters. This is especially true when re-calibrating archival observations to astrometric reference catalogues such as \textit{\textit{Gaia}} which postdate the observations.

In this way, astrophysical missions have not only a scientific purpose but can also obtain a societal spin-off. This paper reports an exploratory pilot for that societal spin-off that consists not only of the re-use of astrophysical imaging data for precovery but also in the re-use of investments put in an information system called AstroWISE to make it generically capable to do precovery in a wide range of astronomical archives within a single data flow environment and with a single precovery pipeline. This exploratory AstroWISE precovery pilot was driven by the following quantitative and qualitative questions:  
\begin{enumerate}
    \item  \textit{What fraction of detectable NEOs can we precover? And how automated can we get the precovery?}
    \item  \textit{What astrometric and photometric accuracy can be achieved?}
    \item \textit{What level of automation can be in achieved the precovery workflow?}
    \item  \textit{What are the major challenges for exploiting various astronomical surveys for NEO space safety purposes?}
    \item  \textit{What value do these precoveries have for planetary defence?}
    \item  \textit{Do precoveries have benefits for NEO science?}
\end{enumerate}
The structure of the paper is as follows. Section 2 describes the data and lists of NEOs used for precovery. In section 3, the pipeline is described in detail. In section 4 the findings and results of NEO precovery are presented and in section 5, we summarize and conclude with the main results of this work and foresee the next steps\footnote{The output of this paper, including FITS files and catalogues, are accessible via the \href{https://www.astro.rug.nl/~neo/}{Kapteyn's NEO data archive}.}.
\begin{figure}
    \centering
    \includegraphics[width=0.99\linewidth]{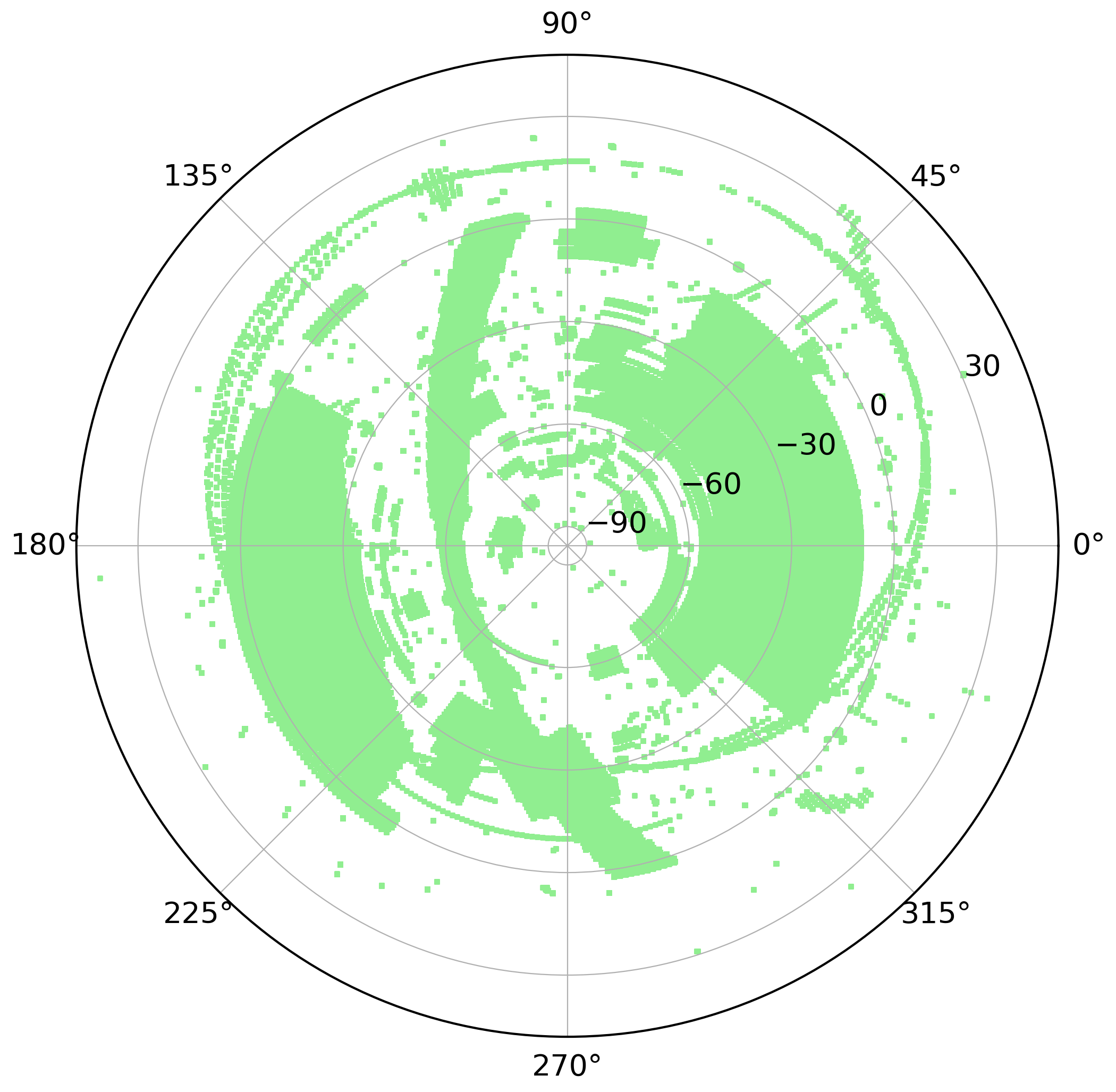}
     \includegraphics[width=0.8\linewidth]{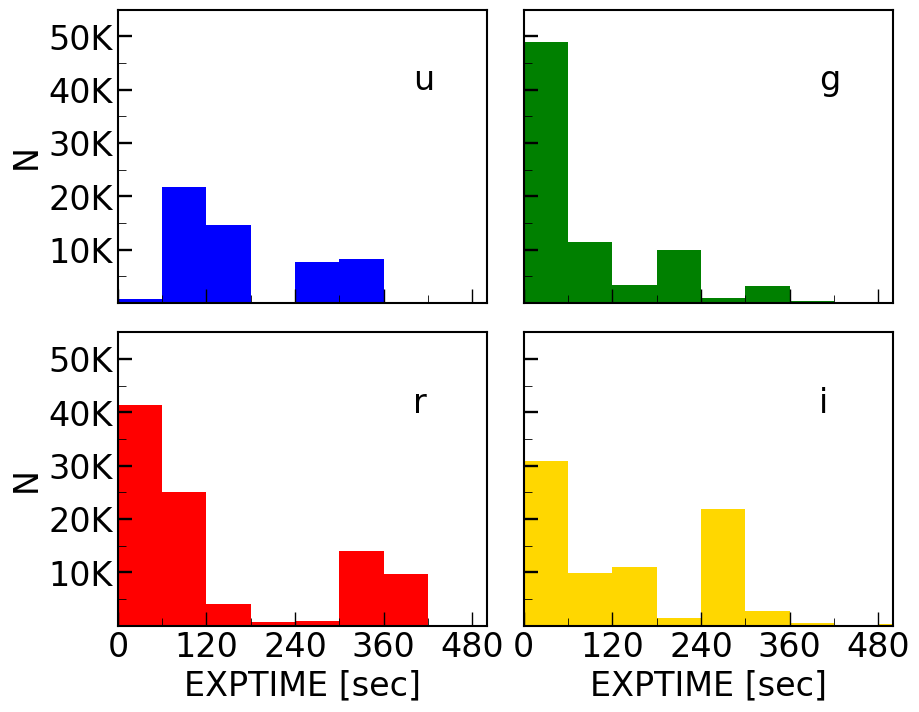}
    \caption{\textit{Top:} the sky coverage of the OmegaCAM/VST observations over 10 years of observations of the southern hemisphere. The \textsc{Astro-WISE} archive contains 361977 individual exposures. \textit{Bottom:} the histogram of exposure times of the OmegaCAM/VST observations in $u$ (54918 exposures), $g$ (79127 exposures), $r$ (96811 exposures) and $i$ (78876 exposures).}
    \label{ocam-coverage}
\end{figure}

\section{Data, software systems \& applications}
In this work, we use the archival imaging data of OmegaCAM, the wide-field imaging camera of the VLT Survey Telescope (VST) at ESO's Cerro Paranal Observatory. In the past OmegaCAM has been used for several wide-field surveys including the KiDS (Kilo-degree Survey, \citealp{kids-dr4}), VST-ATLAS surveys (\citealp{atlas}), the Fornax Deep Survey (FDS, \citealp{fds}) and VEGAS (\citealp{vegas}). OmegaCAM has 32 science CCDs with a Field of View of about 1$^{\degr} $ $\times$ 1$^{\degr} $ on the sky. In its first decade of science operation, OmegaCAM has covered a large fraction of the southern hemisphere (Fig. \ref{ocam-coverage}). Considering the large volume of the dataset of OmegaCAM, we have developed a pipeline to automatize the process of (p)recovery. Here we focus our (p)recovery effort on two datasets of NEOs:
\begin{enumerate}
\item  NEO's \textbf{risk list}\footnote{\url{https://neo.ssa.esa.int/risk-list}}: this list is provided by the ESA near-Earth Objects Coordination Centre (ESA-NEOCC) and consists of about 1350 known NEOs with a non-negligible chance of impact in the next hundred years. This list is updated regularly based on the most recent observations. We use the list of 1 February 2022.
\item  NEOs \textbf{full list}
: this list consists of all the known NEOs, about 30~000 Sources. We used the version available on 1 June 2022.

These data are processed and analysed using several public web-tools to collect the predicted properties of NEOs, using the \textsc{AstroWISE} system (\citealp{aw2,awe}) for data management, image processing, calibration and using dedicated software applications for streak detection, astrometry and photometry. We briefly introduce those software systems and software applications 

\textsc{AstroWISE} stands for Astronomical Wide-field Imaging System for Europe. \textsc{AstroWISE} is an environment consisting of hardware and software which is federated over institutes over Europe. It has been developed to scientifically exploit the ever-increasing avalanche of data produced by astronomical wide-field surveys. \textsc{AstroWISE} is an all-in-one system: it allows a scientist to process raw data, calibrate data, perform post-calibration scientific analysis and archive all the results in one environment. The system architecture links together all these commonly discrete steps in data analysis. 
\end{enumerate}
\section{Precovery Pipeline}
\begin{figure*}
    \centering
    \includegraphics[width=0.9\linewidth]{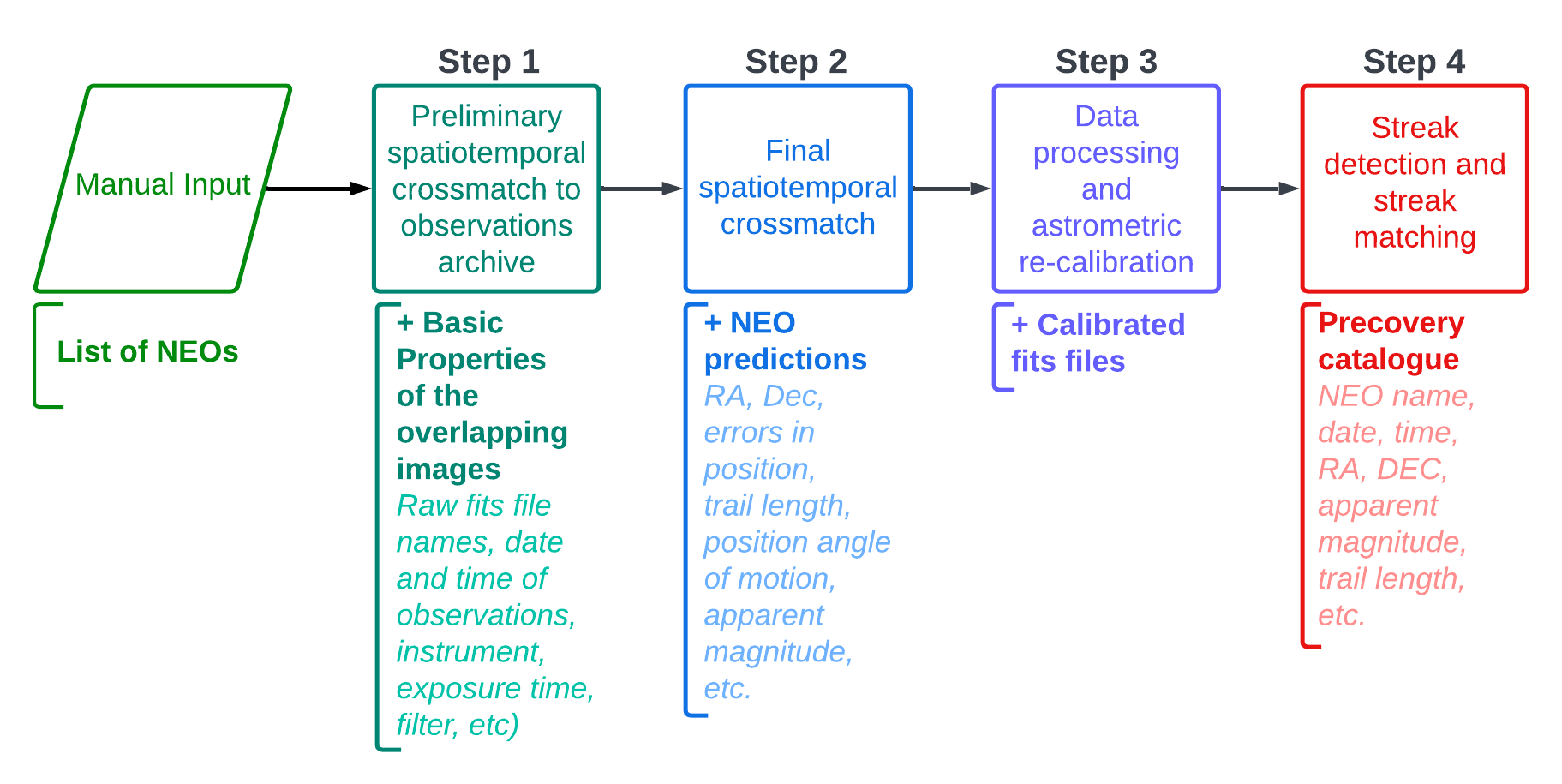}
    \caption{The precovery pipeline. This Diagram shows the four steps in the precovery Pipeline. The pipeline gets a text file containing NEO names as the input and produces a CSV table at the end containing the NEO precoveries. The intermediate outputs are stored in an SQLite database which communicates with the pipeline in many instances throughout the process.}
    \label{pipeline}
\end{figure*}

\begin{figure*}
    \centering
    \includegraphics[width=0.99\linewidth]{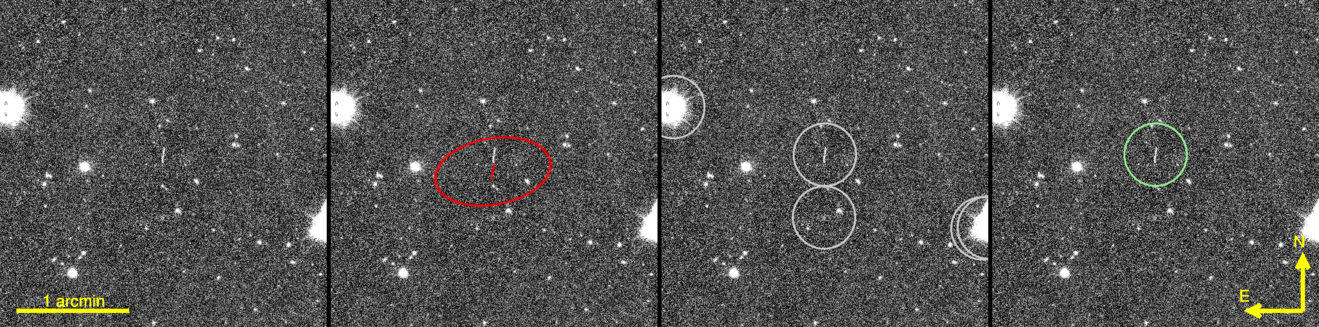}
    \includegraphics[width=0.99\linewidth]{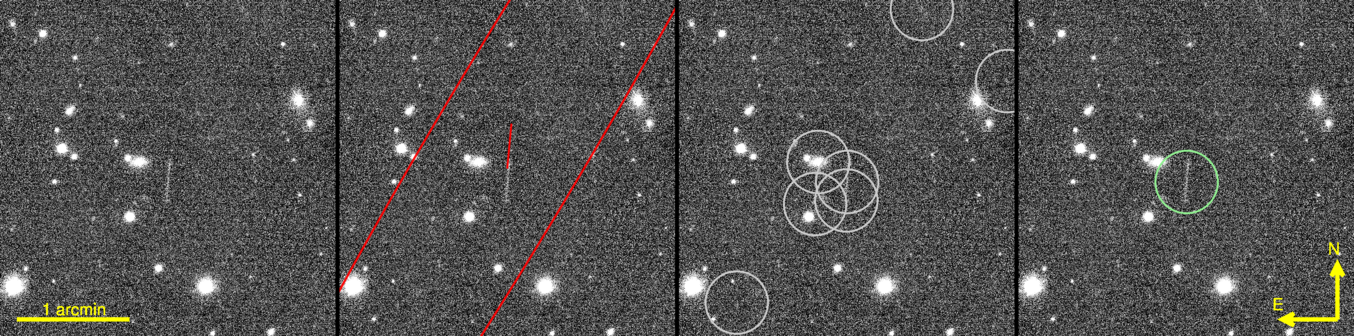}
    \includegraphics[width=0.99\linewidth]{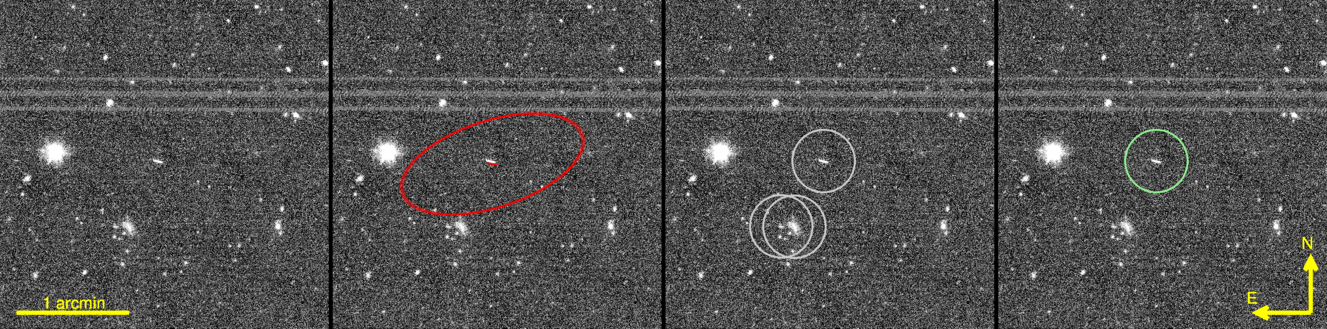}
    \caption{The results of the streak detection using \textsc{StreakDet} for three examples. The first column (from left to right) shows the calibrated cutout around the predicted position of NEOs. In the second column, the 3$\sigma$ uncertainty ellipse, and the predicted appearance of the NEO (considering its speed and direction of motion) are shown (red ellipse and red line). The third column shows the detected streaks across frames by \textsc{StreakDet} (grey circles). Frames in the fourth column indicate the streaks that match the predicted properties of the NEOs (green circles).}
    \label{examples}
\end{figure*} 
The precovery pipeline, shown in Fig. \ref{pipeline}, consists of four main steps. During  these steps, it interacts with several public web-tools to collect the predicted properties of NEOs, and uses the \textsc{AstroWISE} system for data management, image processing, and calibration while using dedicated software applications for streak detection, astrometry and photometry. The four steps are described next.
\subsection{Preliminary spatiotemporal crossmatch to observational archive (step 1)}
As the first step, the possible appearance of NEOs from the provided list  in exposures in ESO’s OmegaCAM archive was assessed using the Solar System Object Image Search (SSOIS)\footnote{\url{https://www.cadc-ccda.hia-iha.nrc-cnrc.gc.ca/en/ssois/}} web service (\citealp{Gwyn}). We used SSOIS as follows: 
\begin{enumerate}

    \item For the NEOs, after a search by object name, the orbital elements provided by the Canadian Astronomy Data Centre (CADC) itself were used, which is a regularly updated copy of the MPC orbital element database. 
    \item SSOIS itself then generated an internal ephemerides database using the \textsc{orbfit}\footnote{\url{http://adams.dm.unipi.it/orbfit/}} software package. 
    \item The SSOIS' internal database of ($RA$, $Dec$) pointings and observation dates for ESO's OmegaCAM observations were used. This database was up to date with observations until approximately 1 November 2021\footnote{private communication with SSOIS custodian Stephen Gwyn}.
    \item The ephemerides database in point 2 was cross-matched in the 3-dimensional space of (observation date, $RA$, $Dec$) by SSOIS itself with the internal database of OmegaCAM observations in point 3. The cross-match algorithm is described in \citet{Gwyn}.
    \item Uncertainties in input orbital parameters and image pointing are not taken into account in the cross-match procedure. This means that the cross-match produces only a result when an NEO ephemeris for the exact orbital elements specified in point 1 falls inside the field of view (FoV) of an OmegaCAM exposure for the exact (observation date, $RA$, $Dec$) specified in point 3.      
    \item  We did not ask SSOIS to refine the cross-match to the ($X$, $Y$) of a specific detector as that functionality is not available (yet) for OmegaCAM. 

\end{enumerate}

To automate our usage of SSOIS we implemented a scripted interface to SSOIS. This interface submits a query to SSOIS. In return, SSOIS provides a list of (raw) imaging data (exposures) for OmegaCAM, date, time, exposure time and the observed filter. SSOIS outputs for OmegaCAM point to the main raw frames (and do not specify which extension (chip) overlaps with the predicted position of the NEO).  The information returned by SSOIS on NEO names and OmegaCAM exposures was stored in a local SQLite file. 

\subsection{Final Spatiotemporal crossmatch (step 2)}
In Step 2, the pipeline uses the Near-Earth Objects Dynamic Site (NEODyS)\footnote{\url{https://newton.spacedys.com/neodys/}} and JPL Horizons\footnote{\url{https://ssd.jpl.nasa.gov/horizons/app.html}} web-service\footnote{In the earlier version of the pipeline, it used NEODyS. In the later versions, we substitute NEODyS with the Horizons system.} to obtain more accurate predictions of the NEO positions over the duration of exposure and to obtain the predicted angular motion and visual magnitude. We supply to NEODyS/Horizons the start and end time of the observation, the name of the NEO and the observatory identifier to specify its location. This query to NEODyS/Horizons returns the ($RA$, $Dec$), 1$\sigma$ uncertainties in position, V$-$band magnitude and the position of the NEO at the beginning and end of the exposure. These values are used to predict the rate and direction of NEO motion relative to the astrometric reference frame. 

Based on the returned apparent magnitude we make a preliminary prediction on the signal-to-noise (SNR) for the given filter and exposure time of the candidate (p)recovery observation by assuming its appearance is a point source (i.e., neglecting proper motion) and assuming a solar spectral energy distribution of the NEO. Candidate (p)recoveries for which NEODyS/Horizons confirm the overlap with the exposure FoV and which have also a predicted SNR\,$>$\,1 are stored in the local SQLite database. To automate our usage of NEODyS/Horizons we implemented a scripted interface to NEODyS/Horizons, using the same approach as for SSOIS.
\subsection{Data Processing and Astrometry (step 3)}
In step 3, we apply several filters and criteria to the precovery candidates resulting from step 2 to narrow down the search to sources which are likely to be detectable. These filters are:
\begin{itemize}
    \item[$\bullet$] an upper limit on the angular separation ($RA$, $Dec$) between SSOIS and NEODyS/Horizons predictions: the imaging data used for precovery are the ones that were identified by SSOIS and based on its predictions. However, once we refine these predictions using NEODyS/Horizons, the images may not overlap with the refined position. Therefore, if the angular separation between these predictions is larger than a certain limit (for example larger than the FoV), we might assume that the object is not within the FoV of the camera.
    \item[$\bullet$] an upper limit on the 1$\sigma$ uncertainty in the position ($RA$, $Dec$): for cases where the uncertainties in position are too large, the possibility that the NEO is within the FoV is very small. The pipeline skips these frames.
    \item[$\bullet$] availability of the raw data: in some cases, the raw frames are not publicly available. The pipeline skips these frames.
    \item[$\bullet$] the exact position of NEOs on the camera: given the coordinates of the object, the pipeline inspects which CCD is overlapping with the corresponding position of NEOs and if there is no overlap, the pipeline excludes these frames.
    \item[$\bullet$] a lower limit on the predicted SNR: this limit will remove cases with an SNR lower than the given limit to exclude very faint and non-detectable objects.
\end{itemize}
All the above-mentioned criteria are applied to both NEO samples i.e. the risk list and the full list. Additionally, for the full list, several other filters are used:
\begin{itemize}
    \item[$\bullet$] a lower limit on the 1$\sigma$ uncertainty in the position: with this filter, we can focus the precovery search on objects without precise positional information. 
    \item[$\bullet$] a lower limit on the predicted length: considering the observational difficulties for detecting and matching short streaks (step 4, described later in the text), the pipeline skips frames with NEOs shorter than this limit.
    \item[$\bullet$] objects without calibrated data: in rare cases, the pipeline fails to reduce/calibrate raw frames and therefore, no calibrated frame is available for further analysis. The pipeline skips these frames.
\end{itemize}
\begin{table}
\centering
\caption{Summary of the filters applied to the NEO candidates in step 3 of the precovery pipeline}
\begin{tabular}{ lcc } \hline \\ Precovery Step & value for &  value fo$r$\\
& risk-list &  full-list\\\\
\hline
\\
Initial SNR cut (in step 2) & 1~ & 1~  \\
Upper limit on trail length & --- & 3\arcsec \\ 
Upper Limit on angular separation & 1$^{\degr}$ & 1$^{\degr}$ \\
Upper limit on  1$\sigma$ errors in position & 1$^{\degr}$ & 1$^{\degr}$ \\
Lower limit on 1$\sigma$ errors in position & --- & 1\arcsec \\
Limit on SNR & 3~ & 3~  \\
\\
\hline 
\end{tabular}
\label{table-filters}
\end{table}
These used thresholds for these filters are expressed in table~\ref{table-filters}. Once a frame passes all the filters and criteria (referred to as a  \textit{precovery candidate}), we then use \textsc{AstroWISE} to produce astrometrically and photometrically calibrated pixel images for the candidate precovery images returned by NEODyS/Horizons. 

If calibrated detector images already exist in the database, they were downloaded from \textsc{AstroWISE}. If not, the raw data was ingested from the ESO archive into the \textsc{AstroWISE} system as needed and the \textsc{AstroWISE} optical image pipeline was used to process that raw data. To obtain optimal astrometry, we manually astrometrically recalibrated the detector images downloaded from \textsc{AstroWISE} using SCAMP 2.10 (\citealp{scamp}). This version of SCAMP uses the coordinates of stars in the \textit{Gaia} EDR3 catalogue as astrometric reference objects. The final astrometry was extracted from the resampled and background subtracted pixels using \textsc{SWarp} (\citealp{swarp}). Note that the majority of OmegaCAM data residing in \textsc{AstroWISE} have been astrometrically calibrated using the near-IR 2MASS catalogue (\citealp{2mass}) as the astrometric reference catalogue. 

To improve the astrometric solution of our dataset and to achieve better astrometry of the precovered NEOs, we implemented astrometric calibration using the \textit{Gaia} EDR3 catalogue (\citealp{gaia-edr3}) as the astrometric reference. It allows us to (re)-calibrate OmegaCAM images astrometrically to \textit{Gaia}. It improves the external astrometric accuracy from a typical RMS$~$0.3\arcsec to RMS$~$0.04\arcsec (a factor 7 improvement). This accuracy in astrometric calibration brings down the RMS of the external astrometric residuals below the size of most observed offsets between predicted and precovered astrometry. 

\begin{figure}
    \centering
    \includegraphics[trim={0 0 0 0},width=\linewidth]{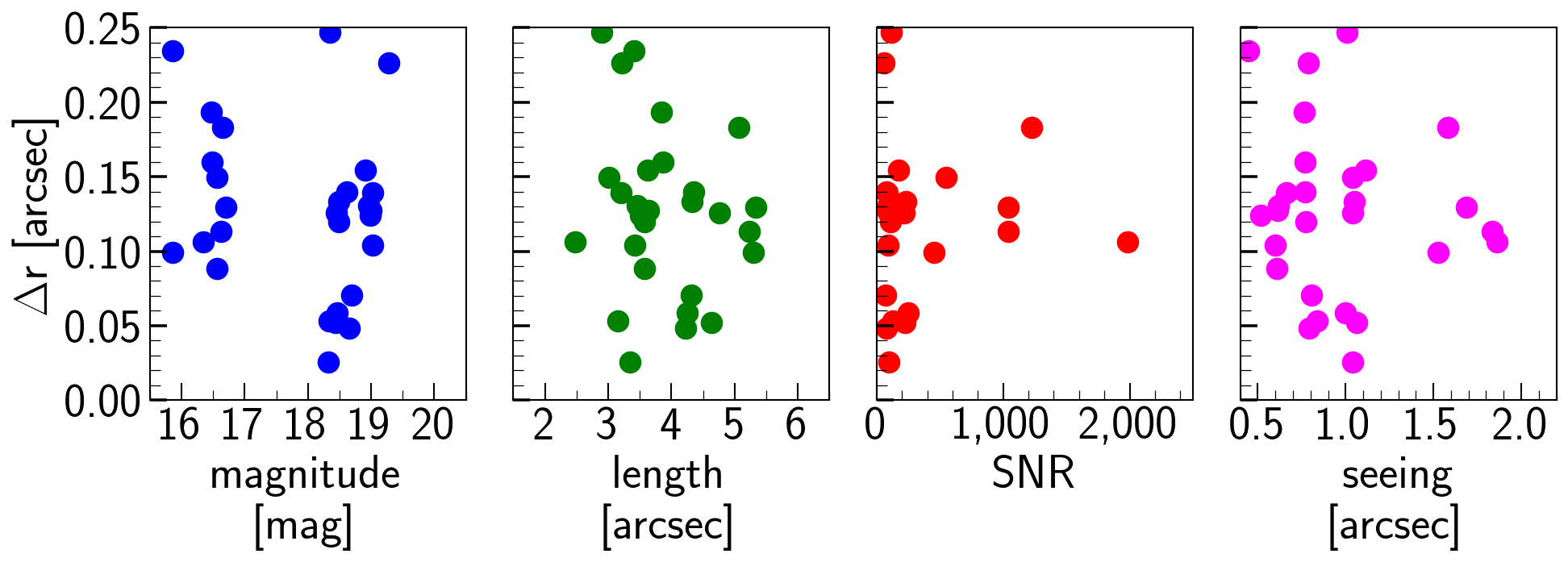}
    \caption{An upper limit for the centroid estimation accuracy of \textsc{StreakDet} in a range of magnitude, length, SNR and seeing. The astrometric accuracy is within 0.12\arcsec and does not show any dependence on the observed properties of NEOs.}
    \label{astrom}
\end{figure} 

\begin{figure}
    \centering
    \includegraphics[width=0.99\linewidth]{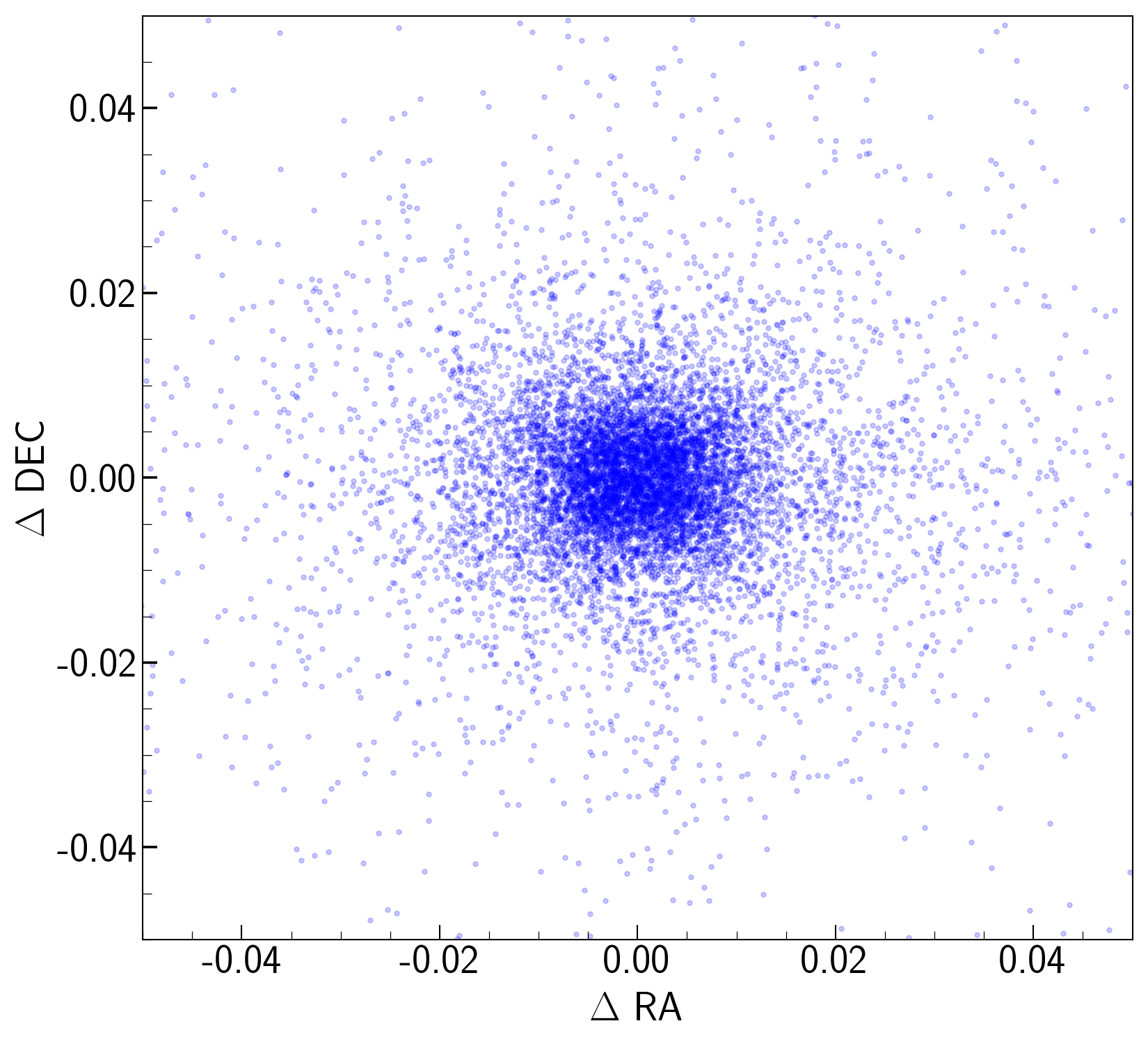}
    \caption{The offset between obtained $RA$ and $Dec$ of all the sources extracted (using \textsc{SExtractor}) in frames with and without resampling and background subtraction (using \textsc{SWarp}) in step 3 of the pipeline}
    \label{astrometry-resampling}
\end{figure} 


\begin{figure}
    \centering
    \includegraphics[width=0.99\linewidth]{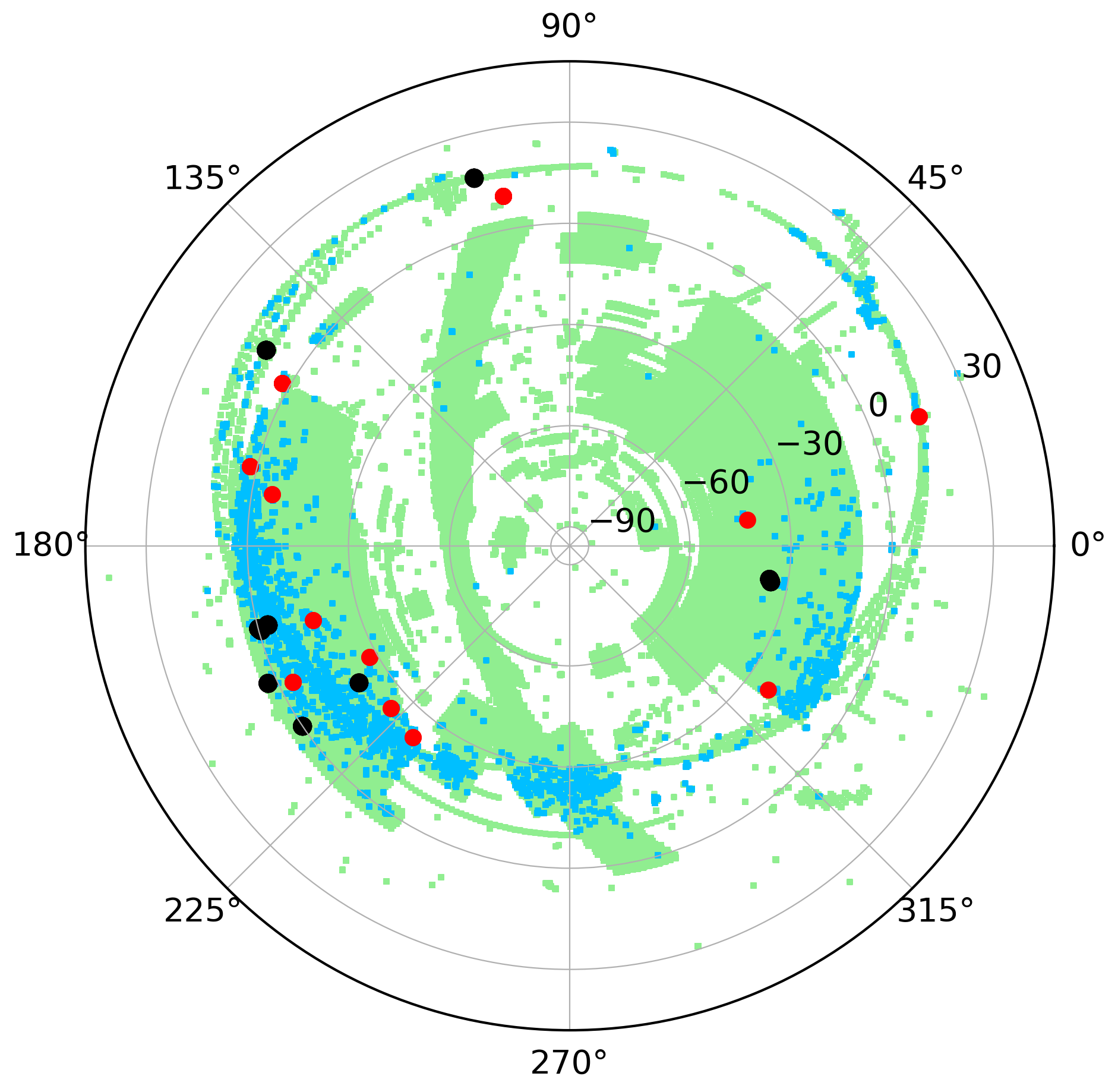}
    \caption{The sky coverage of the OmegaCAM/VST observations (green area), the 10~345 frames resulted with a chance of overlapping with an NEO resulting from SSOIS (blue points), 19 precovery candidates without a detection (red points) and 49  precovered NEOs (black points). Out of these 49 precoveries, 27 are identified using the pipeline, and 22 by visual inspection after step 3 of the pipeline. Note that due to close sky proximity of observations at the resolution of the figure, multiple observations can appear to be a single data point.}
    \label{ocam-coverage-2}
\end{figure} 

\begin{figure*}
    \centering
    \includegraphics[width=0.99\linewidth]{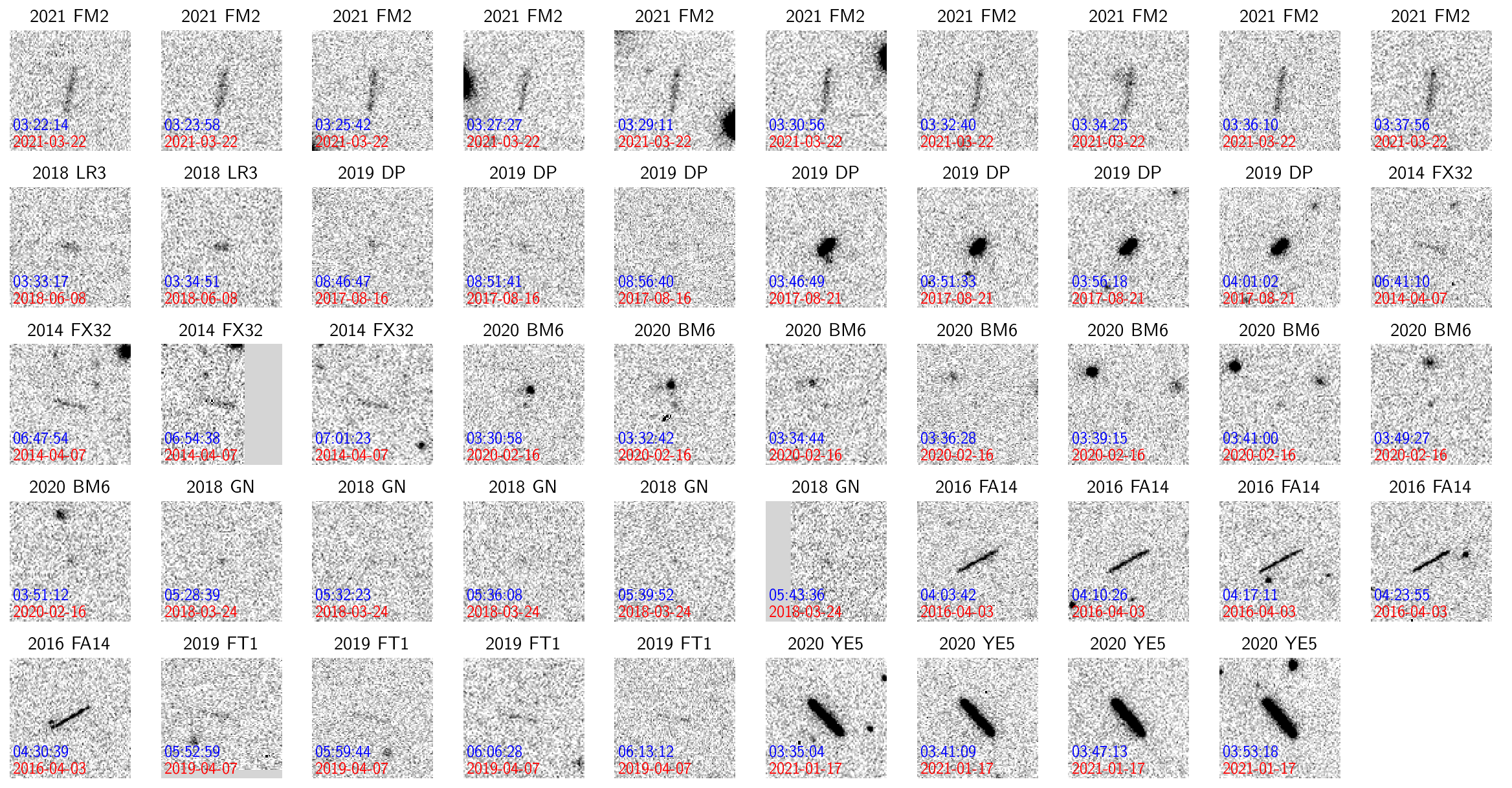}
    \caption{The 49 precovered risk list NEOs using the precovery pipeline. 26 precoveries are detected (fully) automatically and 23 are detected by visual inspection of the frames.}
    \label{risk list-thumbnails}
\end{figure*}

\begin{figure*}
    \centering
    \includegraphics[width=0.48\linewidth]{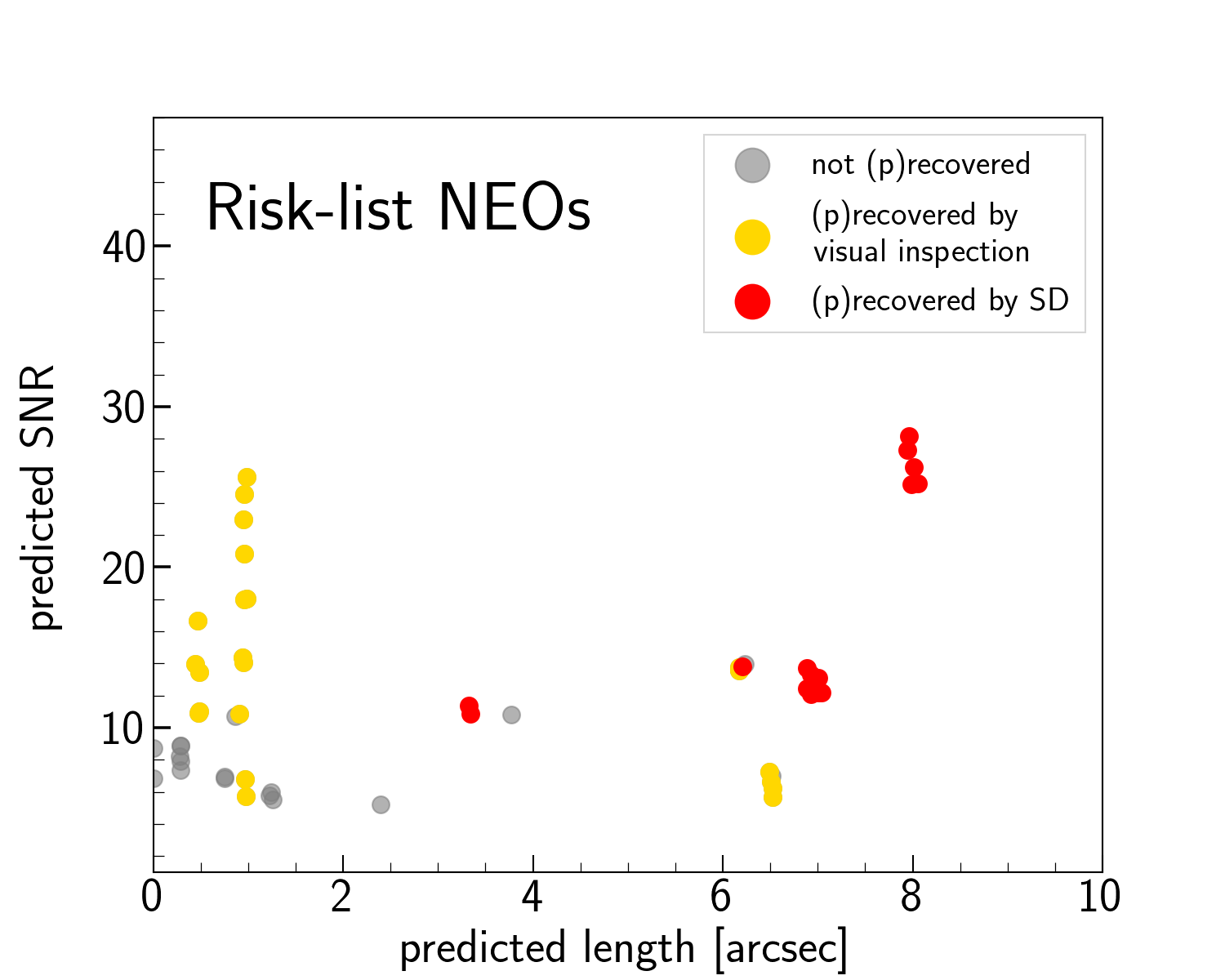}
    \includegraphics[width=0.48\linewidth]{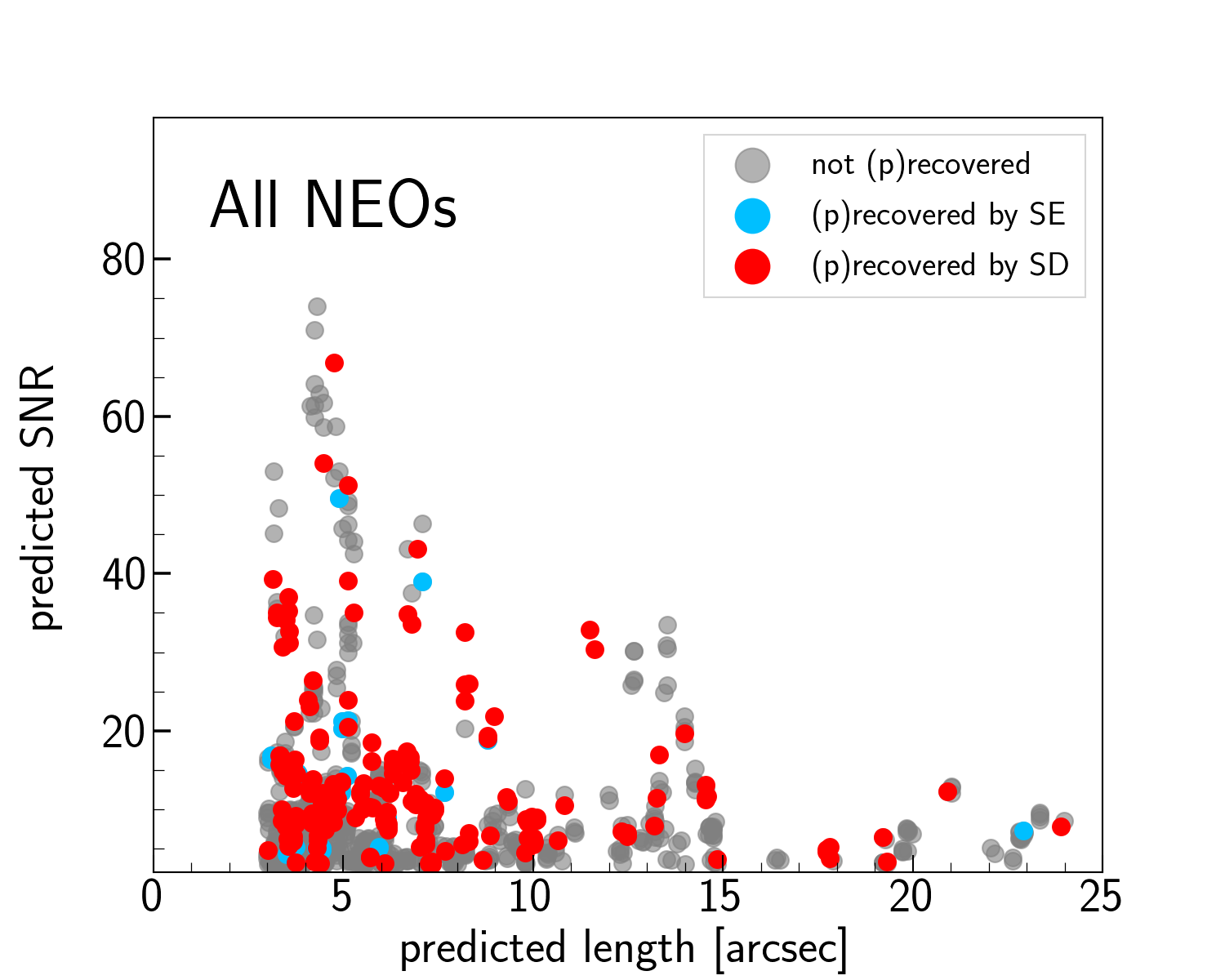}
    \caption{\textit{Left:} SNR and length for precovered NEOs of the risk list $-$ automatic using the pipeline (red points) or manually by visual inspection of frames (yellow points) $-$ and not-precovered risk list NEOs in the OmegaCAM data (grey points). \textit{Right:} SNR and length for precovered NEOs of the full list $-$ using \textsc{StreakDet} (red points) and \textsc{SExtractor} (blue points) $-$ and not-precovered risk list NEOs in the OmegaCAM data (grey points).}
    \label{neos-snr-length}
\end{figure*}

\begin{figure}
    \centering
    \includegraphics[trim={0 0 0 0},width=0.99\linewidth]{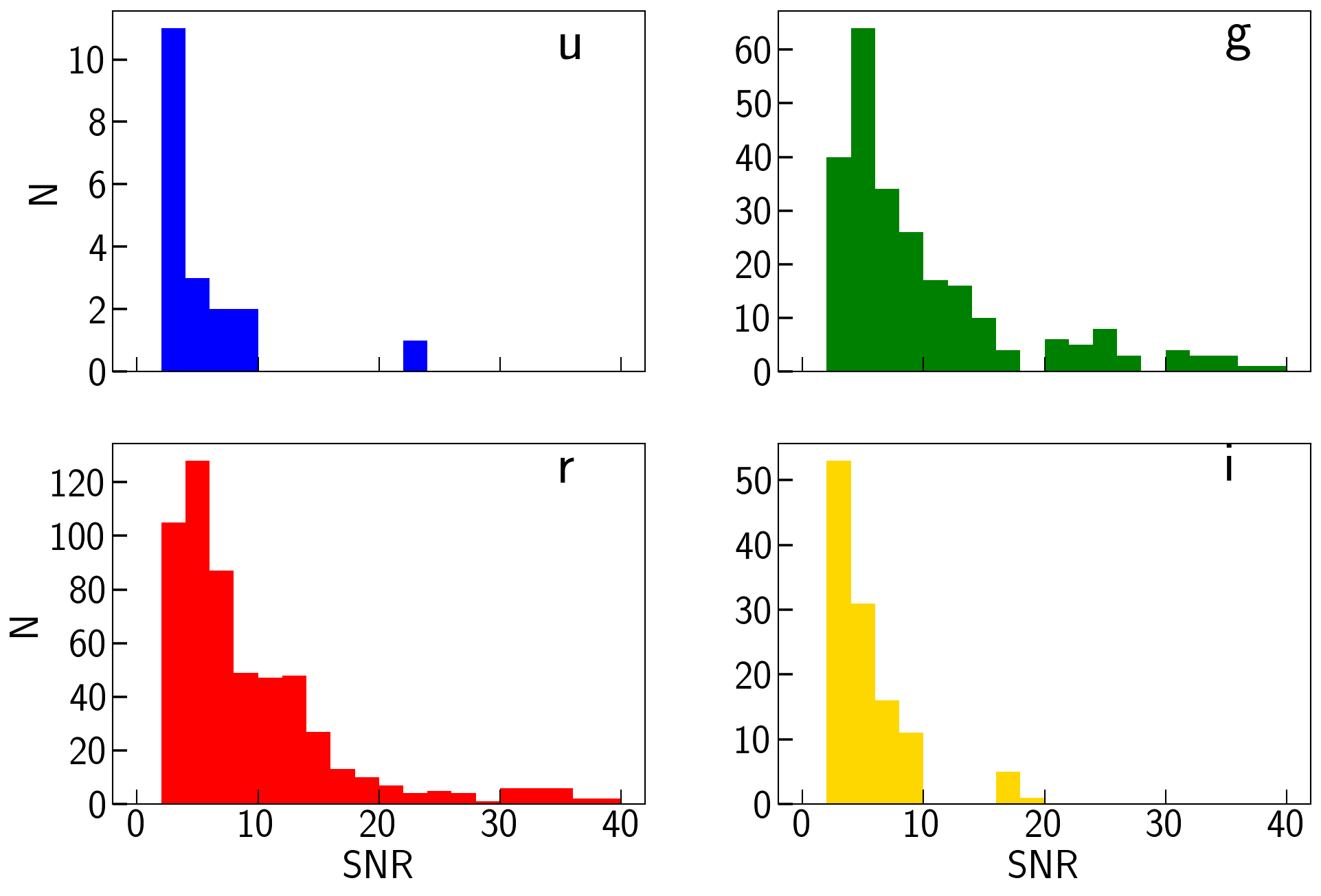}
    \caption{Histogram of the SNR distribution of the 968 precovery candidates after detectability filtering (in step 3) in $u$, $g$, $r$ and $i$.}
    \label{snr-hist}
\end{figure}

\begin{figure}
    \centering
    \includegraphics[trim={0 0 0 0},width=0.99\linewidth]{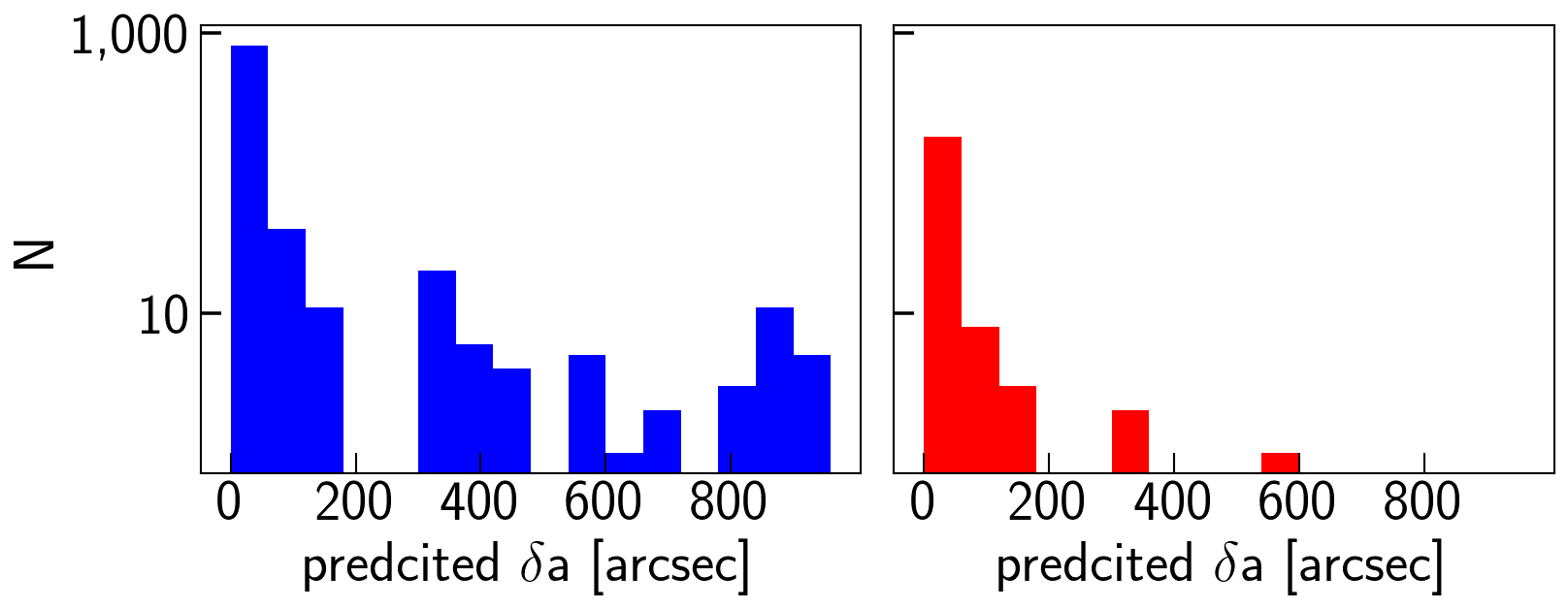}
    \caption{Histogram of the 1$\sigma$ positional uncertainties of the 968 precovery candidates after step 3 (\textit{left}) and the successful precoveries by the pipeline (\textit{right}).}
    \label{da-hist}
\end{figure} 


\begin{figure*}
    \centering
    \includegraphics[width=0.99\linewidth]{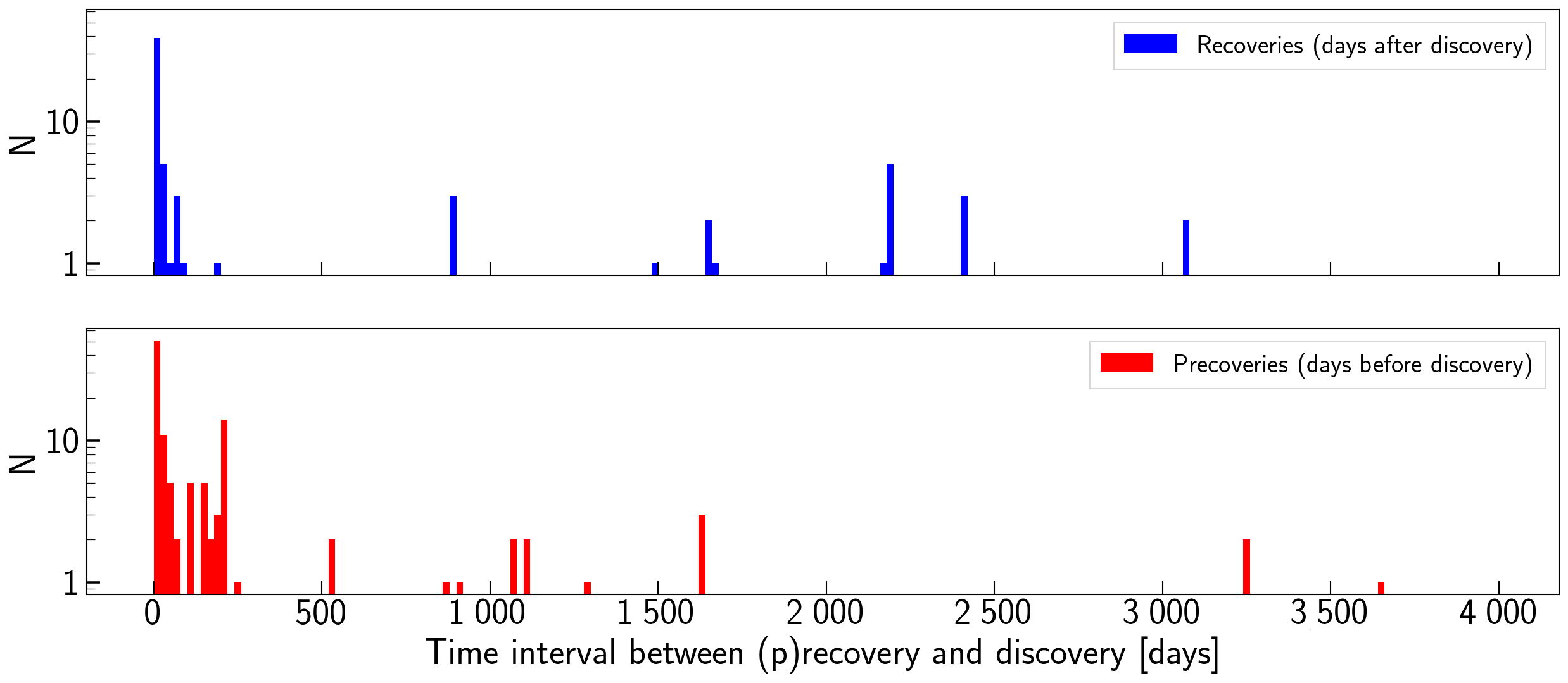}
    \caption{The time interval between precovery and discovery of the 196 NEO appearances (NEO full-list). The upper and lower panels show the NEO appearances after the discovery (recovery) and before the discovery (true precovery) of the NEOs respectively.}
    \label{precovery-date}
\end{figure*}


\begin{figure*}
    \centering
    \includegraphics[width=0.48\linewidth]{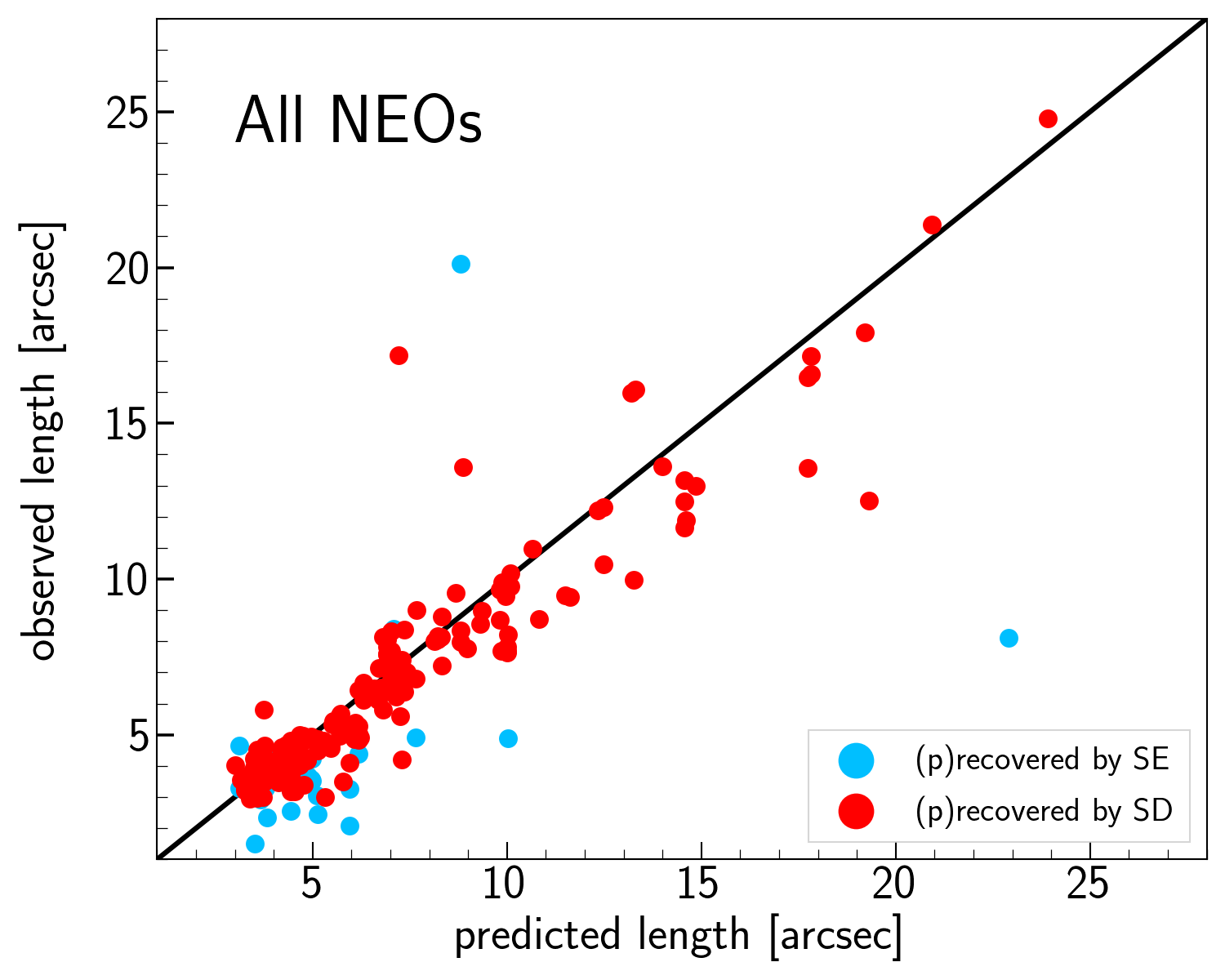}
    \includegraphics[width=0.48\linewidth]{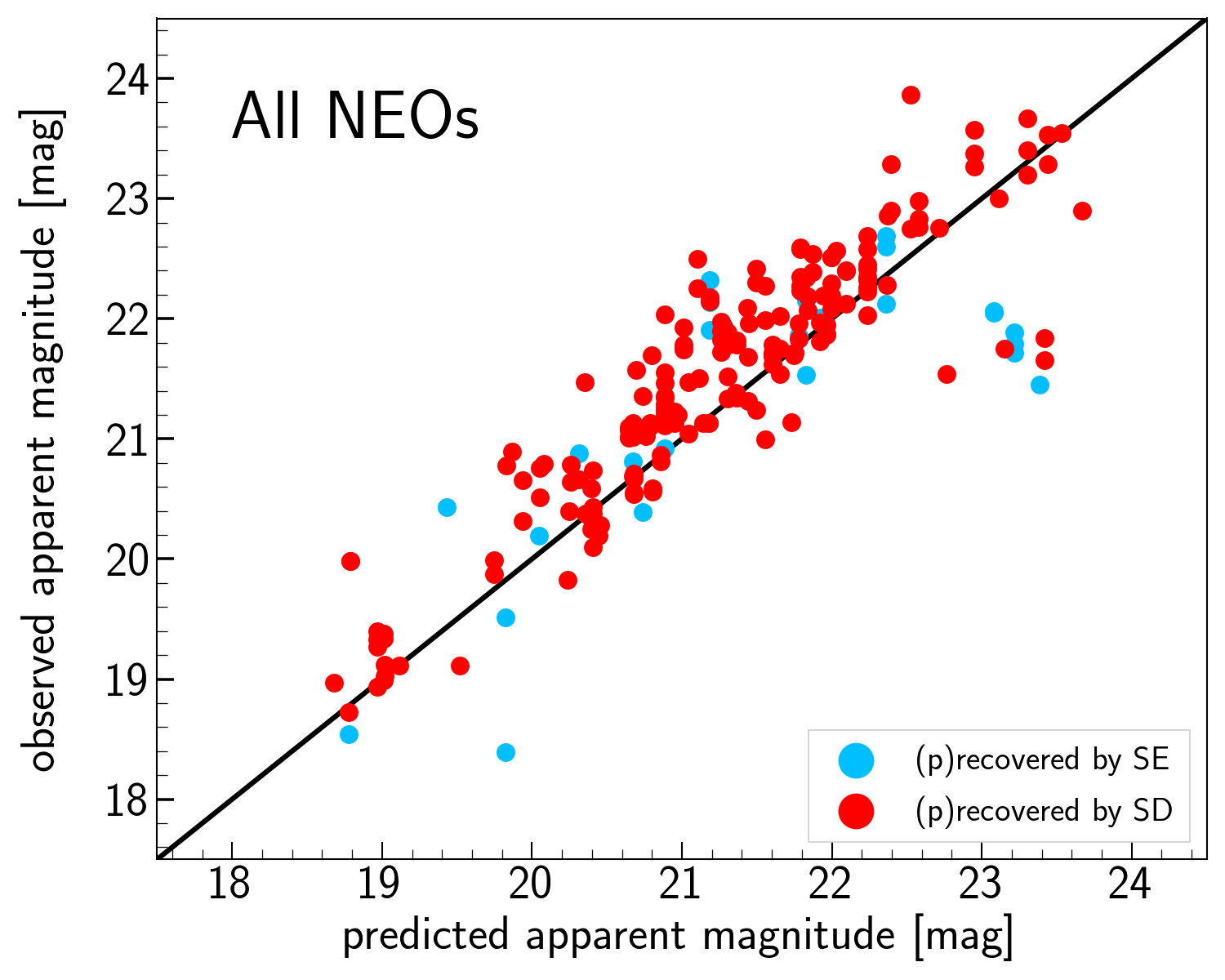}
    \caption{Predicted versus observed length (\textit{left}) and magnitude(\textit{right}) of detected NEOs of the full list by \textsc{StreakDet} (red points) and \textsc{SExtractor} (blue points).}
    \label{snr-length-sex-and-SD}
\end{figure*}

\begin{figure*}
    \centering
    \includegraphics[width=0.48\linewidth]{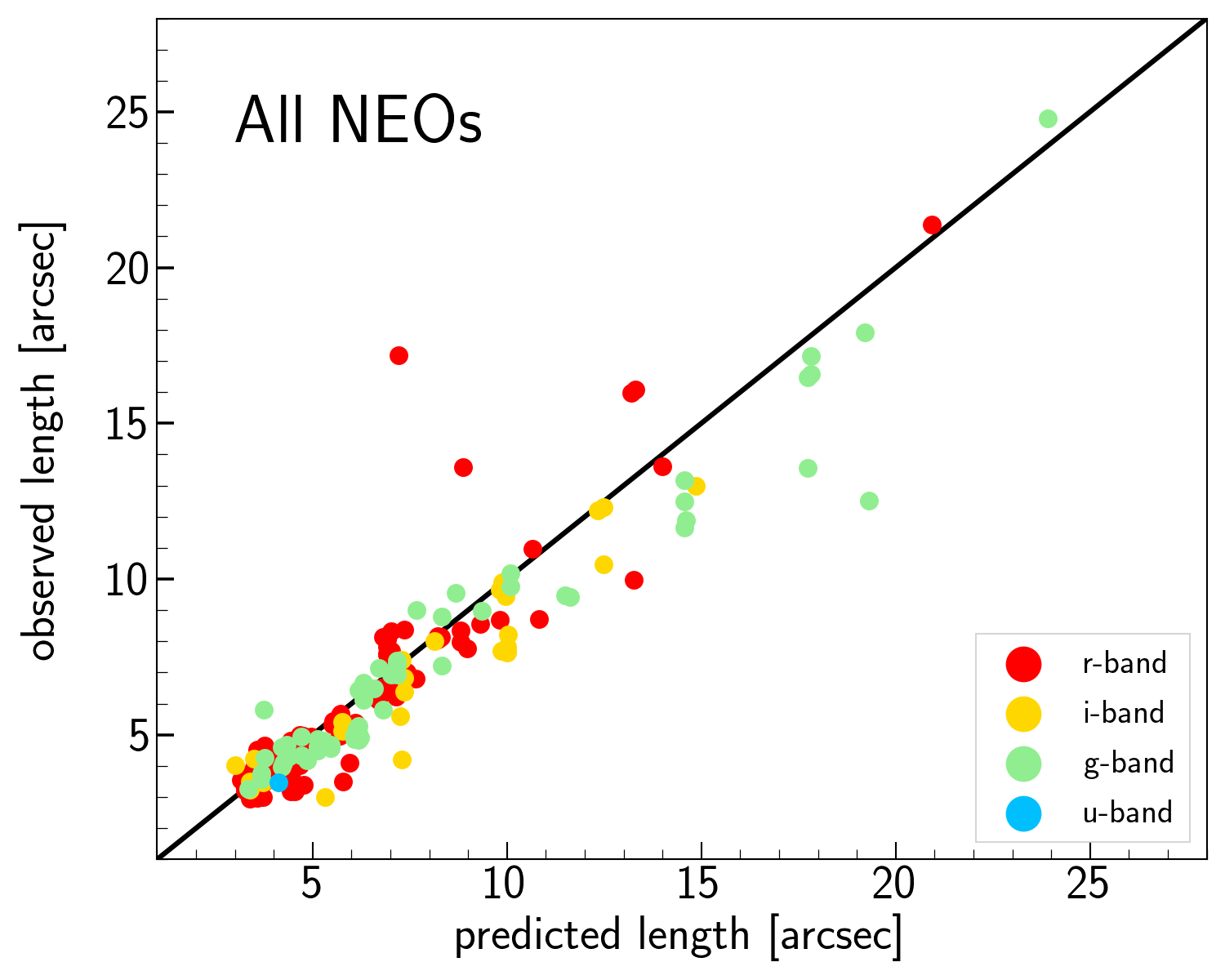}
    \includegraphics[width=0.48\linewidth]{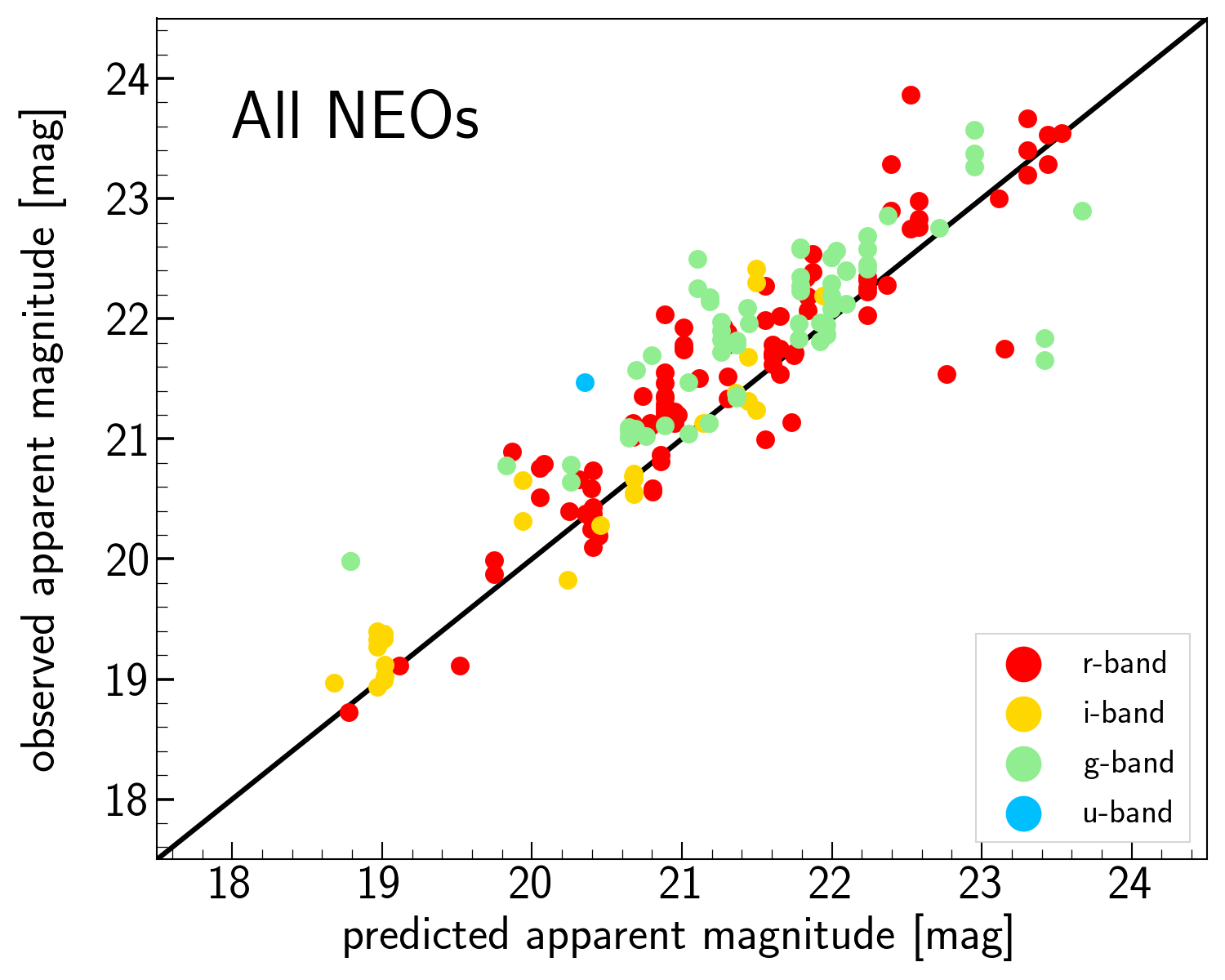}
    \caption{Same as Fig. \ref{snr-length-sex-and-SD} for recoveries done by \textsc{StreakDet} in each filter ($ugri$).}
    \label{snr-length-sex-and-SD-fot-filters}
\end{figure*}

\subsection{Streak Detection and Association (step 4)}
In step 4, the pipeline deploys \textsc{StreakDet} (\citealp{streakdet,streakdet2}) on the photometrically and astrometrically calibrated images to extract candidate streaks. \textsc{StreakDet} is a package that was developed for the purpose of identifying streaks in ground-based and space-based imaging data. \textsc{StreakDet} detects streaks in the images in four steps:  segmentation of images, detection of streak-like objects, classification of objects to separate streaks from other astronomical objects and finally deriving their parameters (such as coordinates, total flux and length) by fitting a model. For the precovery pipeline, we have optimized the \textsc{StreakDet} configuration parameters for OmegaCAM.
Once streak detection is done, then, the pipeline searches for the best match between the streaks found within the 3$\sigma$ uncertainty ellipse of NEOs and their predicted properties. These predicted properties used include the NEOs length, the direction of motion and the magnitude. This procedure is illustrated in Fig. \ref{examples}. The matching procedure (alternatively called "association") is done in two stages. 

In the first step, the pipeline filters out (excludes) the detected streaks with properties different from the predicted properties of the NEOs. The applied filters are:\\
\\$-2$\,mag\,$ < m_{det,\,streak}-m_{pred} < 2$\,mag,
\\... and $0.5 < l_{det,\,streak}/l_{pred} < 2$,
\\... and $0.5 < \sigma_{det,\,streak}/\sigma_{pred} < 1.5$, 
\\... and $-30^{\degr} < \theta_{det,\,streak}-\theta_{pred} < 30^{\degr}$, \\
    ... or $150^{\degr} < \theta_{det,\,streak}-\theta_{pred} < 210^{\degr}$,\\\\
in which $m_{det,\,streak}$, $l_{det,\,streak}$, $\sigma_{det,\,streak}$ and $\theta_{det,\,streak}$ are the observed apparent magnitude (in the given filter), length, width and position angle of motion of the detected streak while $m_{pred}$, $l_{pred}$, $\sigma_{pred}$ and $\theta_{pred}$ are the same properties that are predicted for the NEO.

In the second step, for the detected streaks that satisfy the first step (if any), the pipeline estimates a difference score which is the standardized Euclidean distance between the observed properties of a streak and the predicted properties of the NEO. The parameters used for measuring the distance are the angular length and the position angle of motion, normalized by the observed length and the seeing (FWHM) during observations as follows\footnote{In the earlier version of the pipeline that was applied for precovery of the risk-list NEOs, matching was done based on the angular offset between streaks and the predicted position ($RA$,$Dec$) for NEOs.}:
\\\\
$\theta_{det,\,streak,\,norm} = \theta_{det,\,streak} / Arctan(FWHM/l_{det,\,streak})$\\
$l_{det,\,streak,\,norm} = l_{det,\,streak} / FWHM$
\\\\
in which $\theta_{det,\,streak,\,norm}$ and $l_{det,\,streak,\,norm}$ are the normalised position angle of motion and length of the detected streak. Correspondingly, the difference score is estimated using:\\\\
$score = \sqrt{\theta{_{det,\,streak,\,norm}}^2 + l{_{det,\,streak,\,norm}}^2}$
\\\\
In the end, the detected streak with the lowest difference score is picked as the best match (indicated by the green circle in the fourth column of Fig. \ref{examples}). Finally, those frames with a matched streak are visually inspected to check if it is a true precovery. 
%

\section{Results}

This section presents the outcome of NEO precovery and discusses the performance of the precovery pipeline for the detection, astrometry and photometry of NEOs.
\subsection{Accuracy of astrometry and photometry}
The output catalogues of \textsc{StreakDet} provide the coordinates ($X$,$Y$) of the centre and the total flux of the detected streaks. However, they do not provide uncertainties of these measurements. To make an estimation of the centroiding accuracy of \textsc{StreakDet}, we run the pipeline for NEOs with predicted positional uncertainty of less than 0.02\arcsec (1-$\sigma$ uncertainty ellipse). This is about 10 times smaller than the pixel size of the instrument. The offset between the predicted coordinates of these NEOs ($RA$, $Dec$) and the derived coordinates using \textsc{StreakDet} gives an upper limit of the total astrometric accuracy (i.e., the aggregate of calibration and centroiding accuracies). The average value of this offset, as shown in Fig. \ref{astrom}, is about $\sim$0.12\arcsec.

A preliminary estimate of the astrometric calibration accuracy is $\sim$0.05\arcsec.  It is reasonable to assume that the centroiding and calibration astrometric errors are independent because (i) the centroiding accuracy of the bright \textit{Gaia} stars is likely $<<$~0.05\arcsec and (ii) the same or at least very similar ensemble of \textit{Gaia} stars are used for the astrometric calibration of the exposures for one apparition of an NEO. Note that the absolute astrometric accuracy of stars in the \textit{Gaia} EDR3 is $<<$~0.05\arcsec. Thus we draw the preliminary conclusion that the total absolute astrometric accuracy of our NEO positions is typically $\sim$0.12\arcsec. This means offsets of $>$~0.2\arcsec (i.e., the angular scale of 1 OmegaCAM pixel) between extracted and predicted positions are at typically 1.5$\sigma$ significance prior to taking prediction uncertainties into account in the significance calculation. In individual cases the accuracy can be worse: see notes on individual recoveries below. Moreover, we also assessed the impact of resampling in step 3 of the pipeline (using \textsc{SWarp}) in the astrometry and found a deviation less than 0.01\arcsec (Fig. \ref{astrometry-resampling}). 

In the absence of any indication of photometric accuracy of \textsc{StreakDet}, we made an estimation using the output catalogue of \textsc{Sextractor}. The measurement uncertainty in the measured magnitudes of streaks (within their magnitude range) is on average 0.1\,mag.

\subsection{Precovery of NEOs in the ESA risk list}
Table \ref{table-risklist} summarizes the down selection of precovery candidates during the deployment of this pipeline on the NEOs in ESA’s risk list in combination with ESO’s OmegaCAM archive. This archive encompasses about 10 years of OmegaCAM exposures with a total number of exposures in the order of 400~000.  

The query to SSOIS returned 10~345 exposures, i.e., candidate precoveries for about 1 in 40 exposures. Unfortunately, for 96\% (9904 exposures) of those the upper limit to the predicted SNR of the appearance was $<1$, leaving only 441 candidate precoveries. Subsequently, NEODyS made a refined positional prediction for these 441, which in 33\% of the cases was $>$1$^{\degr}$  away from the centre of the exposure (i.e., the $RA$, $Dec$ of the pointing) which itself is 1$^{\degr}$ $\times$ 1$^{\degr}$. 
For now, we assumed this made it too unlikely that the NEO appearance would be inside the exposure and removed these from the candidate list. This left 295 candidates. For 15\% of those the 1$\sigma$ semi-major axis of the ellipse was $>$1\,$^{\degr}$  and thus again we deemed it too unlikely for now that the NEO appearance would be inside the exposure. This left us with 251 precovery candidate exposures. For 15 of those, the data could not be retrieved in \textsc{AstroWISE}; For 8 it turned out to belong to other OmegaCAM surveys and we are in the process of obtaining access. The remaining 7 are not yet been publicly released by ESO. For 28\% of those the actual predicted position of the NEO landed outside the pixels of the detectors (e.g., in gaps or just outside the FoV) and given the typically small size of the error ellipse we deemed it too unlikely that a precovery could be made. This left 170 precovery candidates. 
At this step, we refined the SNR estimate by taking into account the proper motion. For 60\% of those the predicted SNR was then less than 2. We removed those from the list of candidate precoveries. This left 68 precovery candidates on which precovery was attempted with \textsc{StreakDet}. If \textsc{StreakDet} failed manual precovery was attempted. This was successful for 57\% of the cases, so 49 precoveries (listed in Table \ref{table-of-risklist-precoveries}), with 26 being successful via \textsc{StreakDet} and 23 being successful via subsequent manual analysis. The sky distribution and image thumbnails of these 49 precoveries are shown in Fig. \ref{ocam-coverage-2} and \ref{risk list-thumbnails}. Figure \ref{neos-snr-length} on top shows the predicted properties of detected and non-detected cases. As is seen in this figure, for NEOs with a predicted length smaller than 2\arcsec, we only could detect NEOs by visually examining several frames available. These objects would be missed completely by the pipeline.
%

\subsection{Precovery of NEOs in the full list}
In table \ref{table-all}, we summarize the results from precovery of the full list of NEOs. Queries to SSOIS led to identifying 186~476 frames. In the subsequent step, the query to Horizons removed more than two third of the frames due to the low SNR (SNR$<$1) of the NEOs, which leaves 55~692 cases. Other filters are applied and in each step, a fraction of the precovery candidates are removed. The filters applied to the full list of NEOs are modified slightly to avoid the extra computation time for cases where the pipeline most likely doesn't detect any NEO. For the full list of NEOs, we apply an extra filter on the predicted length of NEOs and do not attempt recovery of objects with predicted lengths smaller than 3\arcsec. These objects, as was seen earlier, are the trickiest to detect (using \textsc{StreakDet}), and they have a similar appearance as other sources (e.g. extragalactic objects and blended objects) which makes the matching process harder. Moreover, a lower limit on the 1$\sigma$ uncertainties in position is introduced to focus our search on objects with a less certain positional prediction. 

After applying all these filters, 968 cases remained, about 0.5\% of the initial number of frames (from SSOIS). Out of 968, after visual inspection of the results, the pipeline successfully detects an NEO in 196 cases (listed in Table \ref{table-of-fulllist-precoveries}), about one-fifth of the total cases. Out of these 196 NEO appearances, 114 are true precoveries. Fig. \ref{precovery-date} shows the histogram of the time intervals for these true precoveries which ranges between 1 day and 3653 days (for 2021 WO), with the majority identified between 1 and 200 days before the discovery date.

The pipeline also results in 157 false positives (about 44$\%$ of automatic precoveries), of which 75$\%$ have a predicted SNR$<$5. Additionally, for long NEOs, the precovery rate decreases because of the limitations in streak detection: \textsc{StreakDet} segments long streaks (low and high SNR) as two separate streaks and the resulting positional information are normally not valid. Therefore in these cases, while the pipeline produces a precovery that almost matches with the NEO, we do not take it into account as a successful precovery.

The bottom graph in Figure \ref{neos-snr-length} shows the predicted SNR and length of the detected and undetected precovery candidates. Out of 772 cases without a precovery, 311 have a predicted SNR$<$5 (Fig. \ref{snr-hist}), and 94 have a 01-$\sigma$ positional uncertainty larger than 200\arcsec (Fig. \ref{da-hist}). The precovery rate of 20$\%$ does not mean that the pipeline can not identify 80$\%$ of the precovery candidates. Visual inspection shows that for the majority of cases without a recovery, no NEO is visible in the frames. 

To examine the possible complementary role of other existing tools for streak detection, we searched for streaks using \textsc{SExtractor} (\citealp{sex}) for the cases without a true detection. \textsc{SExtractor} could identify an extra 26 cases. However, when only  \textsc{SExtractor} is used, the number of detection drops to only 58 (compare to 196 for \textsc{StreakDet}) which shows that \textit{SExtractor} can not substitute \textsc{StreakDet}. Users must be cautious when using \textsc{SExtractor} and consider its limitation to properly segment the streaks which would result in an inaccurate estimate of the centroid of streaks (within a 01-2\arcsec). The uncertainties of \textit{SExtractor} centroiding ($ERRX\_WORLD$ and $ERRY\_WORLD$ parameters in the output tables) is on average 0.88\arcsec, in the same range as the FWHM of the images.
To explore further the accuracy of the predicted properties of NEOs, in figure \ref{snr-length-sex-and-SD}, we compare the predicted and observed (measured from data) length (angular size) and apparent magnitude of the precovered NEOs. There is a general agreement between the predicted and observed properties, in particular for the shorter streaks while the observed NEOs are on average about 0.5\,mag fainter. We explore this magnitude offset further in Fig. \ref{snr-length-sex-and-SD-fot-filters} where the predicted and observed lengths and magnitudes of the precovered NEOs are shown. While there is no clear pattern in the magnitude offset between filters, the predicted and observed magnitudes in $u$ and $g$ show larger offsets. There are two possible sources for this offset: 1. over-prediction of the magnitudes of NEOs, and 2. the simplistic assumption on the colour of NEOs (solar colour) for magnitude transformation between the V-band and the observed filters.
\section{Conclusions and Outlook}
In the introduction section of this paper, we outlined our project objectives and to quantify the results from our exploratory study we defined our main objective to answer six questions. In this section, we answer these questions and add the outlook on possible next steps. 
\begin{enumerate}
    {
    \item  \textit{What fraction of detectable NEOs can we precover? And how automated can we get the precovery?}
    The detectability and the precovery rates vary as a function of the chosen threshold of SNR. The detectability rate is estimated to be $\sim$0.05 per NEO at an SNR larger than 3 for both NEOs on the risk-list and the full list of NEOs. The precovery rate for SNR$>$3 is 40\% for NEOs on the risk-list and 20\% for the full list of NEOs. The precovery rate increases to about 50\% for SNR$>$10. In other words, currently the majority of NEOs with a predicted $3<$SNR$<$10 remain undetected, even after visual inspection. Assessing if the failed precoveries are consistent with predicted errors on the predicted locations and brightnesses is an important next step. This step is beyond the scope of this paper. If inconsistent with predictions it might give new insight on the limitations of those predictions or insights on how to improve the detection/precovery process.  \\
    
    \item  \textit{What astrometric and photometric accuracy can be achieved?}
    The astrometric and photometric accuracies are 0.12\arcsec (15\% of the average FWHM of OmegaCAM/VST images of about 0.8\arcsec) and 0.1\,mag. Improvements in astrometric accuracy are expected from propagating the proper motions in the Gaia astrometric reference catalogue to the observation date of the science image. Improvements in photometric accuracy can come from more sophisticated modelling of SED, observational configuration and NEO shape modelling. The Solar System Open Database Network \citep{berthier22} might facilitate this.\\
    
    \item  \textit{What level of automation can be achieved in the precovery workflow?}
    The precovery pipeline described here works automatically through all the steps. Thanks to the common data model for calibrated observations in \textsc{AstroWISE} \citep{mcfarland13} it can be deployed straightforwardly on calibrated observations for many other instruments available in the \textsc{AstroWISE} archive. However, after the last step (streak detection and matching) and before reporting the recoveries, they must be inspected by an expert for confirming or rejecting the precoveries. Additionally, another challenge is that precise photometric calibration for a range of instruments is hard to fully automate. This is because the derivation of the solution uses reference stars sometimes inside the science images, sometimes in calibration observations. A potential solution would be to construct a photometric reference catalogue that spans the entire sky observable by OmegaCAM with sufficient stellar density. This appears possible by aggregating information from the multiple large-scale surveys of the Southern Sky. 
    \\
    
    \item  \textit{What are the major challenges for exploiting various astronomical surveys for NEO space safety purposes?}
    In addition to the automation already discussed, a main challenge is robust NEO detection and segmentation. This is also a main reason behind the limitations of precovery in this paper. \textsc{StreakDet} is a great tool for detecting high SNR streaks with sizes between 5-20\arcsec. However, its performance drops for faint and long streaks. Deep learning might be a solution to improve streak detection and ultimately NEO precovery. 
    \\
    
    \item  \textit{What value do these precoveries have for planetary defence?}
    The precovery of NEOs provides valuable positional information about NEOs before their discovery. In this paper, we could precover 3 NEOs and as a result, these 3 NEOs are removed from the risk-list of NEOs. 
    \\
    
    \item  \textit{Do recoveries have benefits for NEO science?}
    Precision astrometric information of a sizeable ensemble of recovered NEO can provide a better understanding of non-gravitational effects on NEO orbits (e.g. the Yarkovsky Effect). As a side remark, performing also serendipitous {\it discovery} in astronomical surveys, especially at high ecliptic latitudes, might provide interesting constraints on the orbital demography of NEOs, complementary to NEO dedicated surveys. Collecting the optical/near-infrared spectral energy distribution (SED) information from recovery of a sizeable NEO sample can be complementary to existing astrometric and SED information and hence provide valuable information about their orbit and composition/shapes/rotational properties. 
    }
\end{enumerate}
%
\begin{table*}
\centering
\caption{Summary of ESO OmegaCAM Precovery for the risk list of NEOs.}
\begin{tabular}{ lccc } \hline \\ Precovery Step & Number of &  Number of & Percentage of\\
 & precovery candidates &  precovery candidates & precovery candidates\\ 
 & selected in this step &  removed in this step & removed in this step\\\\
\hline
\\
Query SSOIS & 10~345 & - & - \\
Query NEODyS and an initial SNR cut  & 441 & 9~904 & $\sim$96$\%$ \\
Upper Limit on angular separation limit & 295 & 146 & $\sim$33$\%$ \\
Upper limit on 1$\sigma$ errors in position& 251 & 44 & $\sim$15$\%$ \\
No raw data available on the AW database & 236 & 15 & $\sim$6$\%$ \\
Not covered by the camera (OmegaCAM/VST)& 170 & 66 & $\sim$28$\%$ \\
Limit on SNR & 68 & 102 & $\sim$60$\%$ \\\\
Precovered & 27 & 41 & $\sim$60$\%$ \\
\\
\hline 
\end{tabular}
\label{table-risklist}
\end{table*}

\begin{table*}
\centering
\caption{Summary of ESO OmegaCAM Precovery for the full list of NEOs.}
\begin{tabular}{ lccc } \hline \\ Precovery Step & Number of &  Number of & Percentage of\\
 & precovery candidates &  precovery candidates & precovery candidates\\ 
 & selected in this step &  removed in this step & removed in this step\\\\ 
\hline
\\
Query SSOIS & 186~476 & - & - \\
Query Horizons and an initial SNR cut & 55~692 & 130~514 & $\sim$70$\%$ \\
Lower limit on the predicted length  & 8~440 & 47~252 & $\sim$85$\%$ \\
Upper limit on angular separation limit  & 7~875 & 565 & $\sim$7$\%$ \\
Upper/lower limits on 1$\sigma$ errors in position & 2~683 & 5~192 & $\sim$66$\%$ \\
No raw data available on the AW database & 2~463 & 220 & $\sim$9$\%$ \\
Not covered by the camera (OmegaCAM/VST)& 2~251 & 212 & $\sim$9$\%$ \\
No calibrated data available (calibration failed) & 2~231 & 20 & $<$1$\%$ \\
Limit on SNR & 968 & 1263 & $\sim$57$\%$ \\\\
Precovered & 196 & 772 & $\sim$80$\%$ \\
\\
\hline 
\end{tabular}
\label{table-all}
\end{table*}



\section*{Acknowledgments}
We would like to thank the referee for their comments and suggestions to improve the quality of this work.
We are happy to acknowledge and are grateful to Angela Maria Raj, Katya Frantseva, Migo Müller, Stephen Gwyn, Ylse de Vries, Konrad Kuijken, Danny Boxhoorn, Willem-Jan Vriend, KiDS DR5 production team, Michiel Rodenhuis, Thomas Wijnen, Edwin Valentijn and the Kapteyn institute. This work was executed as part of ESA contract no. 4000134667/21/D/MRP (CARMEN) with their Planetary Defence Office. The Big Data Layer of the Target Field Lab project "Mining Big Data" was used. The Target Field Lab is supported by the Northern Netherlands Alliance (SNN) and is financially supported by the European Regional Development Fund. The data science software system \textsc{AstroWISE} runs on powerful databases and computing clusters at the Donald Smits Center of the University of Groningen and is supported, among other parties, by NOVA (the Dutch Research School for Astronomy). TSR acknowledges funding from the NEO-MAPP project (H2020-EU-2-1-6/870377). This work was (partially) supported by the Spanish MICIN/AEI/10.13039/501100011033 and by "ERDF A way of making Europe" by the “European Union” through grant PID2021-122842OB-C21, and the Institute of Cosmos Sciences University of Barcelona (ICCUB, Unidad de Excelencia ’Mar\'{\i}a de Maeztu’) through grant CEX2019-000918-M. This research has made use of Aladin sky atlas (\citealp{aladin1,aladin2}) developed at CDS, Strasbourg Observatory, France and SAOImageDS9 (\citealp{ds9}). This work has been done using the following software, packages and \textsc{python} libraries: Astro-WISE (\citealp{aw2,awe}), \textsc{Numpy} (\citealp{numpy}), \textsc{Scipy} (\citealp{scipy}), \textsc{Astropy} (\citealp{astropy}).

\bibliographystyle{aa.bst}
\bibliography{main}

\begin{table*}
\centering
\caption{The astrometric and photometric properties of a subset of the 49 precovery cases in this work for risk-list NEOs. Error in magnitudes is 0.1\,mag. The full table is available on the MPC and CDS databases.}
\begin{tabular}{ l c c c c c c } 
\hline \\
NEO name & Date & Time & $RA$ & $Dec$ & magnitude ($mag$) & filter \\\\
2021 FM2 & 2021-03-22 & 03:22:14 & 13:01:08.66 & -00:34:20.71 & 20.5 &  $r$ \\
2021 FM2 & 2021-03-22 & 03:23:58 & 13:01:08.72 & -00:34:08.47 & 20.4 &  $r$ \\
2021 FM2 & 2021-03-22 & 03:25:42 & 13:01:08.78 & -00:33:56.36 & 20.6 &  $r$ \\
2021 FM2 & 2021-03-22 & 03:27:27 & 13:01:08.83 & -00:33:44.17 & 20.9 &  $r$ \\
2021 FM2 & 2021-03-22 & 03:29:11 & 13:01:08.89 & -00:33:32.07 & 20.5 &  $r$ \\
2021 FM2 & 2021-03-22 & 03:30:56 & 13:01:08.94 & -00:33:20.26 & 20.7 &  $r$ \\
2021 FM2 & 2021-03-22 & 03:32:40 & 13:01:08.99 & -00:33:07.93 & 20.8 &  $r$ \\
2021 FM2 & 2021-03-22 & 03:34:25 & 13:01:09.04 & -00:32:56.11 & 20.5 &  $r$ \\
2021 FM2 & 2021-03-22 & 03:36:10 & 13:01:09.09 & -00:32:43.82 & 20.6 &  $r$ \\
2021 FM2 & 2021-03-22 & 03:37:56 & 13:01:09.13 & -00:32:31.86 & 20.7 &  $r$ \\
2018 LR3 & 2018-06-08 & 03:33:17 & 14:12:04.21 & -21:05:29.98 & 22.0 &  $g$ \\
2018 LR3 & 2018-06-08 & 03:34:51 & 14:12:03.87 & -21:05:29.20 & 21.9 &  $g$ \\
2019 DP & 2017-08-16 & 08:46:47 & 23:19:47.23 & -35:09:18.88 & 16.8 & $u$ \\
2019 DP & 2017-08-16 & 08:51:41 & 23:19:47.20 & -35:09:19.17 & 17.0 & $u$ \\
2019 DP & 2017-08-16 & 08:56:40 & 23:19:47.13 & -35:09:19.51 & 17.2 & $u$ \\
2019 DP & 2017-08-21 & 03:46:49 & 23:21:54.59 & -35:34:13.05 & 19.2 &  $i$ \\
2019 DP & 2017-08-21 & 03:51:33 & 23:21:54.42 & -35:34:15.40 & 19.2 &  $i$ \\
2019 DP & 2017-08-21 & 03:56:18 & 23:21:54.24 & -35:34:17.75 & 19.3 &  $i$ \\
2019 DP & 2017-08-21 & 04:01:02 & 23:21:54.06 & -35:34:19.98 & 19.3 &  $i$ \\
... & ... & ... & ... & ... & ... & ...\\
\\
\hline
\end{tabular}
\label{table-of-risklist-precoveries}
\end{table*}
\begin{table*}
\centering
\caption{The astrometric and photometric properties of a subset of the 196 precovery cases in this work for all NEOs. The full table is available on the MPC and the CDS databases.}
\begin{tabular}{ l c c c c c c } 
\hline \\
NEO name & Date & Time & $RA$ & $Dec$ & magnitude ($mag$) & filter \\\\
2011 OV18 & 2012-01-24 & 07:14:13 & 08:33:55.04 & -39:07:50.59 & 21.5 & $u$\\
1997 PN & 2017-05-28 & 09:16:08 & 23:04:46.23 & -27:45:01.63 & 22.2 & $g$\\
1997 PN & 2017-05-28 & 09:19:53 & 23:04:46.58 & -27:44:55.42 & 22.5 & $g$\\
1997 PN & 2017-05-28 & 09:23:38 & 23:04:46.91 & -27:44:49.56 & 22.5 & $g$\\
1997 PN & 2017-05-28 & 09:27:22 & 23:04:47.20 & -27:44:43.80 & 22.1 & $g$\\
1997 PN & 2017-05-28 & 09:31:07 & 23:04:47.53 & -27:44:37.69 & 22.3 & $g$\\
2005 UG1 & 2016-11-04 & 00:52:15 & 00:32:01.77 & -29:03:40.42 & 21.4 & $g$\\
2005 UG1 & 2016-11-04 & 00:56:00 & 00:32:02.60 & -29:03:57.58 & 21.3 & $g$\\
2005 UG1 & 2016-11-04 & 01:03:30 & 00:32:04.42 & -29:04:34.89 & 21.8 & $g$\\
2005 UG1 & 2016-11-04 & 01:07:15 & 00:32:05.34 & -29:04:53.97 & 21.8 & $g$\\
2013 EK28 & 2013-05-20 & 07:00:05 & 15:19:52.35 & +00:59:07.89 & 22.4 & $g$\\
2013 EK28 & 2013-05-20 & 07:03:49 & 15:19:52.54 & +00:59:04.25 & 22.1 & $g$\\
2013 EK28 & 2013-05-20 & 07:07:34 & 15:19:52.75 & +00:59:00.84 & 22.4 & $g$\\
2013 HO & 2013-04-03 & 02:22:16 & 11:19:49.98 & -63:14:32.95 & 21.1 & $g$\\
2013 NJ10 & 2019-07-04 & 05:20:16 & 18:05:24.90 & -24:36:19.65 & 22.2 & $g$\\
2013 NJ10 & 2019-07-04 & 05:26:00 & 18:05:24.50 & -24:36:20.40 & 22.5 & $g$\\
2013 NJ10 & 2019-07-06 & 04:56:46 & 18:02:16.25 & -24:42:00.33 & 22.1 & $g$\\
2013 NJ10 & 2019-07-06 & 05:20:31 & 18:02:14.66 & -24:42:02.90 & 22.1 & $g$\\
2014 HT190 & 2014-04-21 & 02:49:55 & 14:10:24.49 & -01:53:34.15 & 22.9 & $g$\\
... & ... & ... & ... & ... & ... & ...\\
\\
\hline
\end{tabular}
\label{table-of-fulllist-precoveries}
\end{table*}

\end{document}